\documentclass[
aps,%
12pt,%
final,%
notitlepage,%
oneside,%
onecolumn,%
nobibnotes,%
nofootinbib,%
groupedaddress,%
noshowpacs,%
centertags]%
{revtex4}
\usepackage [cp866]{inputenc}
\usepackage [english]{babel}
\usepackage {caption2}
\usepackage{epstopdf}

\def\bge{\begin{equation}}
\def\ene{\end{equation}}
\def\bgea{\begin{eqnarray}}
\def\enea{\end{eqnarray}}
\def\nn{\nonumber}

\def\sfrac#1#2{{#1}/{#2}}

\begin{document}
\selectlanguage{english}
%%%

\begin{center}
{\Large \bf Polarization effects in radiative decay of polarized
$\tau$ lepton
 }
\vspace{4 mm}

G.I.~Gakh$^{(a, b)}$, M.I.~Konchatnij$^{(a)}$, A.Yu.~Korchin$^{(a,
b)}$, N.P.~ Merenkov$^{(a, b)}$

\vspace{4 mm}
%$^{(a)}$Jefferson Lab, Newport News, VA 23606, USA\\
%$^{(b)}$Duke University,
%Durham, NC 27708, USA \\
%$^{(c)}$ Center of Particle and High Energy Physics,
%220040 Minsk, Belarus \\

$^{(a)}$ NSC ''Kharkov Institute of Physics and Technology'',\\
Akademicheskaya, 1, 61108 Kharkov, Ukraine \\
$^{(b)}$ V.N.~Karazin Kharkiv National University, 61022 Kharkov,
Ukraine
\end{center}

\vspace{0.5cm}

\begin{abstract}
 The polarization effects in the one-meson
radiative decay of the polarized $\tau$ lepton,
$\tau^-\to\pi^-\gamma\nu_{\tau}$, are investigated. The inner
bremsstrahlung and structural amplitudes are taken into account.
The asymmetry of the differential decay width caused by the $\tau$
lepton polarization and the Stokes parameters of the emitted
photon itself, are calculated depending on the polarization of the decaying
$\tau$ lepton. The numerical estimation of these
physical quantities was done for arbitrary direction of the $\tau$
lepton polarization 3-vector in the rest frame. The vector and
axial-vector form factors describing the structure-dependent part
of the decay amplitude are determined using the chiral effective
theory with resonances (R$\chi$T).
\end{abstract}

\maketitle

\section{Introduction}
\hspace{0.7cm}

As it is known, the $\tau $ lepton is the only existing lepton
which can decay, due to its large mass, into final states
containing hadrons. The energy region of these decays corresponds
to the hadron dynamics which is described by the non-perturbative
QCD. Since the complete theory of non-perturbative QCD is absent
at present, the phenomena in this energy region are described
using various phenomenological approaches. To test different
theoretical models it is important to investigate experimentally
the hadronization processes of the weak currents. The semileptonic
$\tau $ lepton decays are very suitable for such investigations
since the leptonic weak interaction is well understood in the
Standard Model (SM). A review of the present status of $\tau $
physics can be found in Ref. \cite{P13}.

In the last decade the experimental investigations of the $\tau $
lepton decays are strongly enlarged due to the construction of the
B-factories with very high luminosity (BaBar, Belle)
$L\approx 10^{34}$ cm$^{-2}$ s$^{-1}$ \cite{Z11}. At present the
experiments at the B-factories led to the accumulation of the data
sets of more than $10^9$ $\tau $ lepton pairs \cite{K08}.
Interesting results obtained at the B-factories revived the plans
on the construction of the new facilities such as SuperKEKB
(Japan) and Super $c-\tau $ (Russia) \cite{Z11, Lev08, O09}. These
projects will use a new technique to collide the electron-positron
beams which permits to increase the existing luminosity by one or
two orders of magnitude. The designed luminosity is $L \approx
(1-2) \times 10^{35}$ cm$^{-2}$ s$^{-1}$ for the Super $c-\tau $
and $L \approx 10^{36}$ cm$^{-2}$ s$^{-1}$ for the SuperKEKB
\cite{Z11}. Besides, the Super $c-\tau $ and SuperKEKB factories
can have a longitudinally polarized electron beam with the
polarization degree of more than $80 \%$ and this guarantees
production of polarized $\tau $ leptons.

This very high luminosity of the planned $\tau $ factories will
allow one to perform the precise measurements of various decays of
the $\tau $ lepton and thus permits to search for the
manifestation of the new physics beyond the SM, such as a search
for the lepton flavor violation, $CP/T$ violation in the lepton
sector and so on.

The simplest semileptonic $\tau $ lepton decay is
$\tau^-\to\pi^-(K^-)\nu_{\tau}$, however in this case the
hadronization of the weak currents is described by the form
factors at fixed value of the momentum transfer squared $t$ ($t$
is the difference of the $\tau^-$ and $\nu_{\tau}$ 4-momenta
squared). The dependence of the form factors on this variable can
be determined, in principle, in the transition $W\to \pi(K)\gamma
$, where $t$ is the squared invariant mass of the $\pi (K) -
\gamma $ system. This transition can be investigated in
the $\tau$ lepton radiative decay
$\tau^-\to\pi^-(K^-)\nu_{\tau}\gamma .$

The one-pseudoscalar meson radiative $\tau $ lepton decays have
been investigated in a number of papers \cite{Lev08,O09,K80,B86,I90,D93,R95}.
%\cite{K80 - G10}.
The authors of Ref. \cite{K80} obtained the expression for the
double-differential decay rate for the
$\tau^-\to\nu_{\tau}\pi^-\gamma $ decay in terms of the vector
$v(t)$ and axial-vector $a(t)$ form factors. The numerical
estimates were done for the real parameterizations of these form
factors using the vector-meson dominance approach. But the assumption
that the form factors are real functions is not generally true
since in the time-like region of the momentum transfer squared,
which is the case for considered decay, these form factors are
complex functions.

The author of Ref. \cite{B86} has studied the following radiative
decays $\tau^-\to\nu_{\tau}\pi^-\gamma $ and
$\tau^-\to\nu_{\tau}\rho^-\gamma,$ obtained the analytical
formulas for the differential decay rates and evaluated them
assuming that the form factors are constant. The authors of Ref.
\cite{D93} have studied the decays
$\tau^-\to\nu_{\tau}\pi^-(K^-)\gamma $. They obtained the photon
energy spectrum, the meson-photon invariant mass distribution, and
the integrated rates. The inner bremsstrahlung contribution to the
decay rate contains infrared divergences and that is why the
integrated decay rates must depend on the photon energy cut-off (or
meson-photon invariant mass cut-off). For the photon energy cut-off 100
MeV, the integrated decay rates
$R=\Gamma (\tau\to\nu_{\tau}\pi\gamma )/\Gamma
(\tau\to\nu_{\tau}\pi )=1.4\cdot 10^{-2}$ \cite{K80} and
$R=1.0\cdot 10^{-2}$ \cite{D93} were obtained. Note that the leptonic radiative
decay of the $\tau $ lepton
$\tau^-\to\mu^-\bar\nu_{\mu}\nu_{\tau}\gamma $ was measured with a
branching ratio of $3.6\cdot 10^{-3}$ \cite{PDG}. So one can
expect that the one-pseudoscalar meson radiative $\tau $ lepton
decay $\tau^-\to\nu_{\tau}\pi^-\gamma $ can also be measured
experimentally since theory predicts for its branching ratio
the value of the same order as for the
$\tau^-\to\mu^-\bar\nu_{\mu}\nu_{\tau}\gamma $ decay.

Some polarization observables in the decay
$\tau^-\to\nu_{\tau}\pi^-\gamma $ have been considered in Ref.
\cite{R95}. The general expressions for the Stokes parameters of
the produced photon have been calculated. The influence of the
possible anomalous magnetic moment of the $\tau$ lepton and
existence of excited neutrinos on the matrix element of this
decay are briefly discussed. The authors showed that a measurement
of the dependence of the differential decay rate on the photon
energy (at a fixed sum of the photon and pion energies) allows to
determine the moduli and phases of the form factors as functions
of the variable $t.$

The $\tau$ lepton radiative decays
$\tau^-\to\nu_{\tau}\pi^-(K^-)\gamma $ were also studied in Refs.
\cite{G03, G10} for the case of unpolarized particles. The light
front quark model was used to evaluate the form factors $v(t)$ and
$a(t)$ describing the structure-dependent contribution to these
decays \cite{G03} and they found that in the SM the decay width
is $\Gamma~=~1.62\cdot 10^{-2}(3.86\cdot
10^{-3})\Gamma(\tau\to\nu\pi)$ for the photon energy cut-off of 50 (400)
MeV. The same decays were studied in Ref. \cite{G10}. The photon
energy spectrum, pion-photon invariant mass distribution and
integrated decay rate were calculated without free parameters and
the authors obtained for the decay width values $1.46\cdot
10^{-2}(2.7\cdot 10^{-3})\Gamma(\tau\to\nu\pi)$ for the same cut-off
conditions.

The $\tau $ lepton decay for the case of the virtual photon which
converted into the lepton-antilepton pair has been investigated in
Refs. \cite{Guevara:2013wwa}. This decay was not measured up to now, but
the cross channel (namely, $\pi^+\to e^+\nu_ee^+e^-$) has been
already measured \cite{E89, B92}. This decay and decay $\tau\to
\pi ll\nu_{\tau}$ probe the transition $W^*\to\pi\gamma^*$ where
both bosons (W and photon) are virtual. These decays complement
the decay we consider in this paper and which can be a source of
information about the transition $W^*\to\pi\gamma $. The vector
and axial-vector form factors are the functions of two variables
(instead of one as in our case) due to the virtuality of the
photon and a third form factor appears in this case. The authors
\cite{Guevara:2013wwa} calculated the branching ratios and di-lepton
invariant mass spectra. They predicted that the process with $l=e$
should be measured soon at B-factories.

Since the SuperKEKB and Super $c-\tau $ factories plan to have the
longitudinally polarized electron beam with the polarization
degree of the beam about $80\%,$ it is worthwhile to investigate
the polarization of the $\tau$ lepton.
%Another reason to do this is related to the fact that the
%measurement of the asymmetries caused by the $\tau$ lepton
%polarization is simpler than the measurement of the Stokes
%parameters of the high energy photon.
In this paper we investigated the polarization effects in the
one-meson radiative decay of the $\tau$ lepton,
$\tau^-\to\pi^-\gamma\nu_{\tau}$. The decay polarization asymmetry
and the Stokes parameters of the emitted photon itself have been
calculated for the case of the polarized $\tau$ lepton . Numerical
estimates of these observables were done for arbitrary
polarization of the $\tau$ lepton.

The vector and axial-vector form factors (which are of theoretical
and experimental interest), describing the structure-dependent
part of the decay amplitude, are determined in the framework of
the chiral effective theory with resonances
(R$\chi$T)~\cite{EckerNP321, EckerPL223}. The R$\chi$T is an
extension of the Chiral Perturbation Theory to the region of
energies around 1 GeV, which includes explicitly the meson
resonances. The corresponding Lagrangian contains a few free
parameters, or coupling constants, and at the same time has a good
predictive power.  This approach has further theoretical
developments, such as in Refs.~\cite{Cirigliano_2006,Rosell_2007},
and applications to various aspects of meson phenomenology (see,
{\it e.g.}, review~\cite{Portoles_2010}). Here we would like to
mention earlier studies of $e^+ e^-$ annihilation to the pair of
pseudoscalar mesons with final-state
radiation~\cite{Dubinsky_2005}, radiative decays with light scalar
mesons~\cite{Ivashyn_2008}, two-photon form factors of the $\pi^0,
\, \eta$ and $\eta^\prime$ mesons~\cite{Czyz_2012}, and three-pion decay of the
$\tau^-$ lepton \cite{Nugent:2013hxa}.

The paper is organized as follows. In Sec.~2 the matrix element of
the decay $\tau^- \to \nu_\tau \pi^- \gamma$ is considered, and the
phase space factor of the final particles is introduced for an unpolarized and
polarized $\tau$ lepton, Stokes parameters and spin-correlation
parameters of the photon are defined. It is done with help of the
current tensor $T^{\mu\nu}$ and two unit space-like orthogonal
4-vectors which describe the polarization states of photon and
which we express via particle 4-momenta. In Sec.~3 the current
tensor is calculated in terms of the particle 4-momenta and the
$\tau$ lepton polarization 4-vector. In Sec.~4 we describe the chosen
model for the vector and axial-vector form factors which enter the
structural part of the decay amplitude. In Sec.~5 results of some
analytical and numerical calculations are presented and
illustrated. Sec.~6 contains a discussion and conclusion. In Appendix
A the formalism of R$\chi$T is briefly reviewed.  In Appendix B we
consider a polarization of the $\tau^-$ lepton in the annihilation
process $e^+ e^- \to \tau^+ \tau^- $ near threshold for a
longitudinally polarized electron.

\section{General formalism}
The main goal of our study is the investigation of
polarization effects in the radiative semileptonic decay of a
polarized $\tau$ lepton
\begin{equation}\label{1}
{\tau}\,^-(p)\rightarrow
\nu_{\tau}(p\,')+\pi^-(q)+{\gamma}\,(k)\,.
\end{equation}

\subsection{Amplitude and decay width}

The corresponding Feynman diagrams for the decay amplitude are
shown in Fig.~1. The pole diagrams (a) and (b) describe the
inner bremsstrahlung (IB) by the charged particles in the point-like
approximation; diagram (c) describes the so-called structural
radiation.

\vspace{0.4cm}

%\begin{minipage}{165 mm}
\begin{figure}
%\begin{center}
\captionstyle{flushleft}
\includegraphics[width=0.253\textwidth]{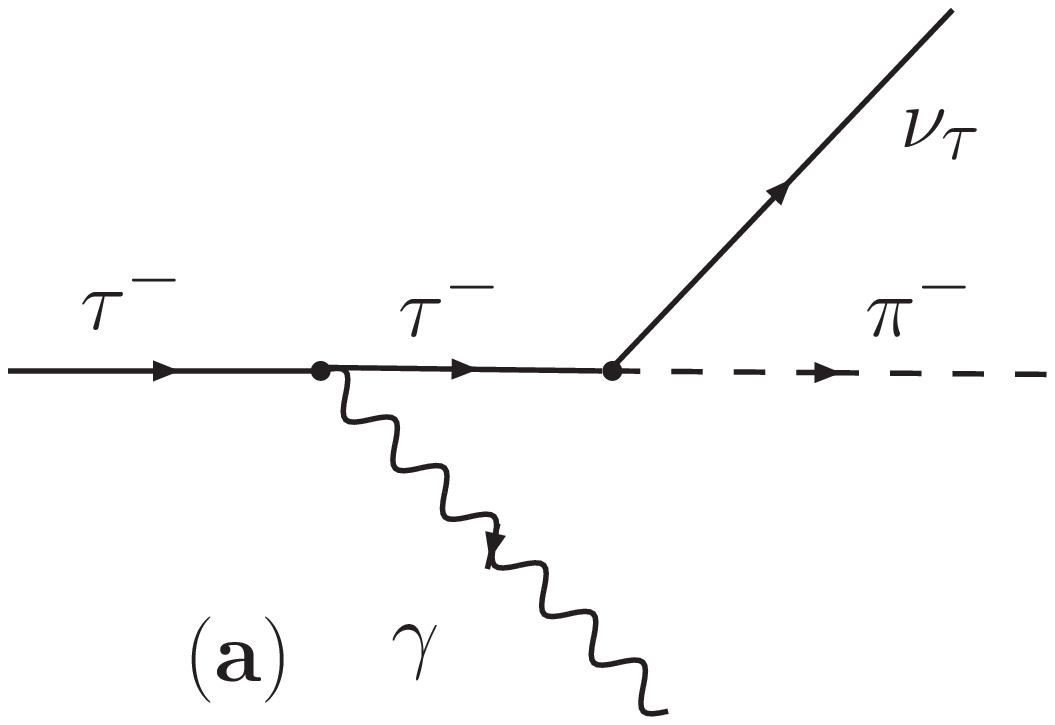}
\hspace{0.5cm}
\includegraphics[width=0.253\textwidth]{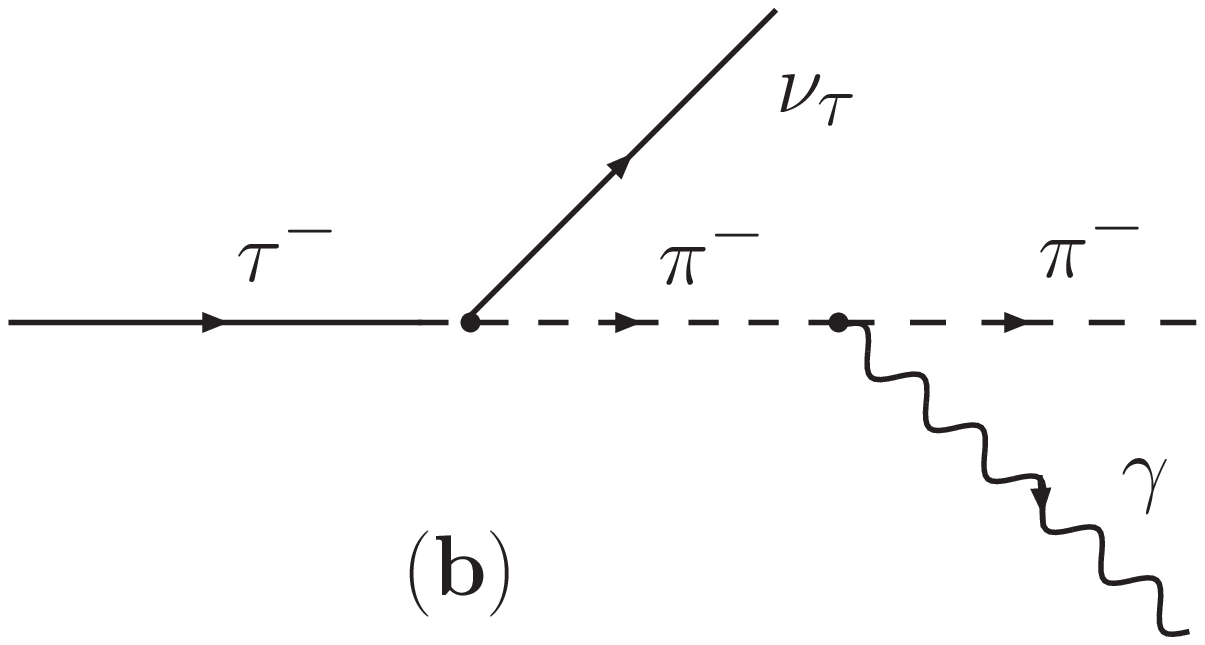}
\hspace{0.5cm}
\includegraphics[width=0.253\textwidth]{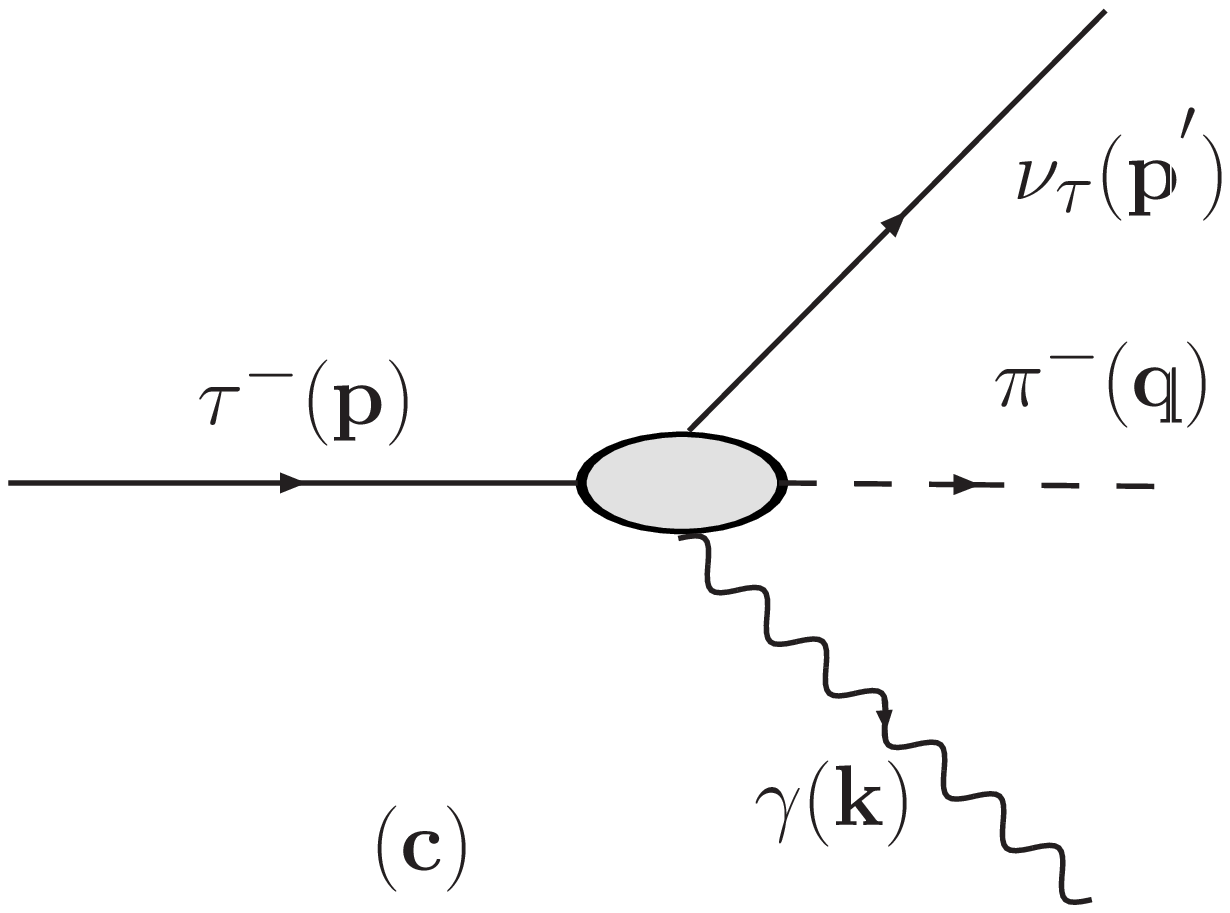}
% \end{center}

%\emph{\textbf{Fig.1.}}
%{\emph{
\caption{ The Feynman diagrams for the
radiative $\tau^-\rightarrow \pi^-+\nu_{\tau}+\gamma$ decay. The
diagrams (a) and (b) correspond to the so-called  structure-independent
inner bremsstrahlung for which it is assumed that the
pion is a point-like particle. The diagram (c) represents the
contribution of the structure-dependent part and it is
parameterized in terms of the vector and axial-vector form
factors. }
\end{figure}
%\end{minipage}

\vspace{0.4cm}

Thus, we have \cite{D93,RekaloLuca}
$$ M_{\gamma}=M_{IB}+M_S \,.$$

The IB piece, in the case of real photon ($k^2=0$), coincides with
its so-called "contact limit" value and reads
\begin{equation}\label{2}
iM_{IB}=ZM\bar u(p')(1+\gamma_5)\Big[\frac{\hat
k\gamma^{\mu}}{2(kp)}+\frac{Ne_1^{\mu}}{(kp)(kq)}\Big]u(p)\varepsilon^*_{\mu}(k)
\ ,
\end{equation}
where the dimensional factor $Z$ incorporates all constants:
$Z=eG_FV_{ud}F_{\pi}$, M is the $\tau$ lepton mass and
$\varepsilon_{\mu}(k)$ is the photon polarization 4-vector. Here
$e^2/4\pi=\alpha=1/137\,,$ $G_F=1\,.166\cdot 10^{-5}GeV^{-2}$ is
the Fermi constant of the weak interactions, $V_{ud}=0.9742$ is
the corresponding element of the CKM-matrix \cite{CKM},
$F_{\pi}=92\,.42 MeV$ is the constant which determines the decay
$\pi^-\rightarrow\mu^-\bar\nu_{\mu}.$ The remaining notation is
$$e_1^{\mu}=\frac{1}{N}\big[(pk)q^{\mu}-(qk)p^{\mu}\big]\ , \ \
(e_1k)=0\ , \ e_1^2=-1\ , $$
$$ N^2=2(qp)(pk)(qk)-M^2(qk)^2-m^2(pk)^2\ ,$$
where $m$ is the pion mass.

The structure radiation in the decay (1) arises due to the possibility
of the virtual radiative transition
$$W^-\rightarrow \pi^- +\gamma.$$
We write the corresponding amplitude in a standard form in terms of two
complex form factors: vector $\, v(t)\,$ and axial-vector $\,
a(t)\,$ \cite{G10,RekaloLuca}
\begin{equation}\label{3}
iM_R=\frac{Z}{M^2}\bar
u(p')(1+\gamma_5)\Big\{i\gamma_{\alpha}(\alpha\mu
kq)v(t)-\big[\gamma^{\mu}(qk)-q^{\mu}\hat k
\big]a(t)\Big\}u(p)\varepsilon^*_{\mu}(k)\ ,
\end{equation}
where $t=(k+q)^2$,  and
$$(\alpha\mu kq) = \epsilon^{\alpha\mu\nu\rho}k_{\nu}q_{\rho}\ ,
\ \epsilon^{0123}=+1\ , \ \
\gamma_5=i\gamma^0\gamma^1\gamma^2\gamma^3\,, \
Tr\gamma_5\gamma^{\mu}\gamma^{\nu}\gamma^{\rho}\gamma^{\lambda}=-4i\epsilon^{\mu\nu\rho\lambda}\
. $$

One can see that both matrix elements, $M_{IB}$ and $M_{R}$, satisfy the condition of gauge invariance and for $M_{R}$
it is valid for any choice of form factors.

The form factors play an important role in the low-energy hadron
phenomenology. However, the experimental values of $v(0)$ and
$a(0)$ have uncertainties both in absolute values and signs
\cite{PDG}. Of course, such a situation  complicates a search for
the signals of new physics beyond SM in future experiments with
high statistic at $\tau$-factories \cite{Exptau, Bl09}.

To define the vector and axial-vector form factors we use the
model based on the Resonance Chiral Theory \cite{EckerNP321} . The
brief physical description of the theoretical approach to this problem
is given in Appendix A and Section 4. In accordance with the
results of the theoretical model used, we can write the form
factors in the following form
$$ a(t)=-f_A(t)\frac{M^2}{\sqrt{2}mF_{\pi}}\ ,  \ \ v(t)=-f_V(t)\frac{M^2}{\sqrt{2}mF_{\pi}}\ ,$$
where $f_A(t)$ and $f_V(t)$ are defined in Section 4.
%by {\bf
%Eqs.(B.10) and (C.5)}, respectively.

We use such a normalization that the differential width of the decay
(1), in terms of the matrix element $M_{\gamma},$ has the following
form in the $\tau$ lepton rest system
\begin{equation}\label{4}
d\Gamma=\frac{1}{4M(2\pi)^5}|M_{\gamma}|^2\frac{d^3k}{2\omega}\frac{d^3q}{2\epsilon}\delta(p'^{2})\
,
\end{equation}
where $\omega $ and $\epsilon$ are the energies of the photon and
$\pi$ meson. When writing $|M_{\gamma}|^2 $ one has to use
$$ u(p)\bar u(p)= (\hat p +M)\ , \ \ u(p)\bar u(p)= (\hat p
+M)(1+\gamma_5\hat S) $$ for unpolarized and polarized $\tau$
lepton decays, respectively. Here $S$ is the 4-vector of $\tau$
lepton polarization.

\subsection{Phase space factor}

It is convenient to analyze events of $\tau$ decay  in its rest
frame. In this system, in an unpolarized case, $|M_{\gamma}|^2$ depends
on the photon and pion energies only: $\omega$ and $\epsilon.$
Then the phase space factor for unpolarized $\tau$ can be written
as \cite{R95}
\begin{equation}\label{5}
d\Phi=\frac{d^3k}{2\omega}\frac{d^3q}{2\epsilon}\delta(p'^{2})\
=\pi^2d\omega d\epsilon\ ,
\end{equation}
and the region of variation of the energies is defined by the
following inequalities

\begin{equation}\label{6}
\frac{M^2+m^2-2M\epsilon}{2(M-\epsilon+|\bf q|)} \leq\omega \leq
\frac{M^2+m^2-2M\epsilon}{2(M-\epsilon-|\bf q|)}\ , \ \ m\leq
\epsilon \leq \frac{M^2+m^2}{2M}\,, \end{equation}
$$\frac{M^2+m^2+4\omega(\omega-M)}{2(M-2\omega)}\leq \epsilon \leq \frac{M^2+m^2}{2M}\,, \ 0\leq\omega\leq \frac{M^2-m^2}{2M}.$$

As the $\tau$ radiative decay amplitude depends on the invariant
variable $t=(k+q)^2=M(2\epsilon+2\omega-M)$ via the vector and
axial-vector form factors in the amplitude (3), one can perform
integration of the differential width with respect to $\epsilon$
(or $\omega$) at fixed values of $t$ to investigate these form
factors. It may be done noting that
$$d\epsilon d\omega~=~\frac{1}{2M}d\epsilon
dt~=~\frac{1}{2M}d\omega dt$$ and
\begin{equation}\label{7}
\frac{t^2+m^2M^2}{2Mt}\leq ~ \epsilon ~ \leq ~
\frac{M^2+m^2}{2M}\,; \ \frac{t-m^2}{2M}~\leq ~ \omega ~ \leq ~
\frac{M(t-m^2)}{2t}\,; \ \ m^2 ~ \leq ~ t ~ \leq ~ M^2\,.
\end{equation}
 The
integration regions for the variables $(\epsilon\,,~\omega)\,, \
(\epsilon\,,~ t)$ and $(\omega\,,~ t)$ are shown in Fig.~2.

In a polarized case we have an additional independent 4-vector $S.$ In
the $\tau$ lepton rest frame $S=(0,\bf n).$ We define a coordinate
system with an axis OZ directed along the vector $\bf n$ and the pion
3-momentum lies in the plane XZ, as it is shown in Fig.~3. If one
used the $\delta$ - function $\delta(p'^2)$ to eliminate integration
over the azimuthal angle $\varphi,$ the phase space (5) can be
rewritten in the form
\begin{equation}\label{8}
d\Phi=\frac{\pi}{2}\frac{dc_1dc_2d\omega d\epsilon}{K}\ , \
K=s_1s_2|\sin\varphi|\sqrt{1-c_1^2-c_2^2-c_{12}^2+2c_1c_2c_{12}}\
,
\end{equation}
where $c_1\,(s_1)=\cos\theta_1\,(\sin\theta_1) , \
c_2\,(s_2)=\cos\theta_2\, (\sin\theta_2), c_{12}=\cos\theta_{12}\
, $ and $\theta_1\ , \theta_2\ , \theta_{12} $ are the angles
between $\bf n$ and $\bf q$, \ $\bf n$ and $\bf k$, \ $\bf q$ and
$\bf k$, respectively. In this case we can study the
spin-dependent effects caused only by the terms, in the matrix
element squared, which do not depend on the azimuthal angle,
namely $(Sk)$ and $(Sq).$  The contribution of the term containing
$(Spqk)$ vanishes when we integrate over $\varphi$ in the whole
region $(0\,, 2\pi).$

\vspace{0.4cm}

%\begin{minipage}{165 mm}
%\begin{center}
\begin{figure}
\captionstyle{flushleft}
\includegraphics[width=0.35\textwidth]{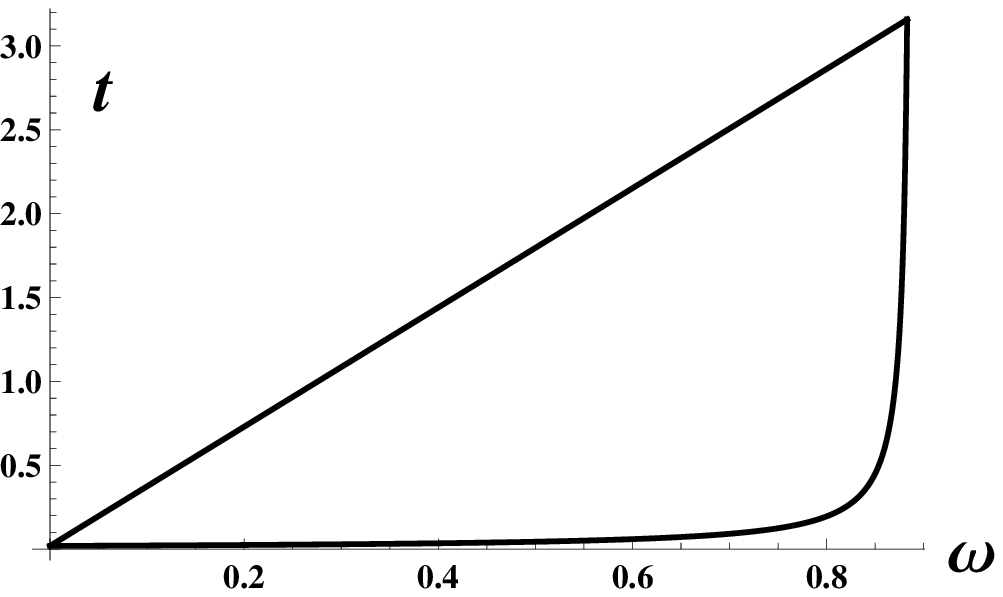}
\hspace{0.5cm}
\includegraphics[width=0.35\textwidth]{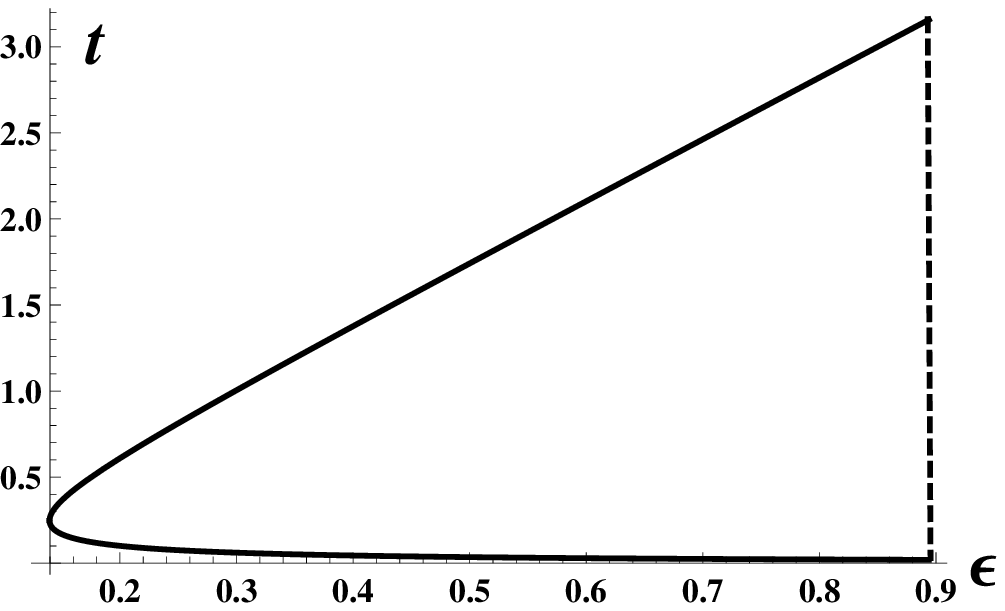}

\vspace{0.5cm}
\includegraphics[width=0.35\textwidth]{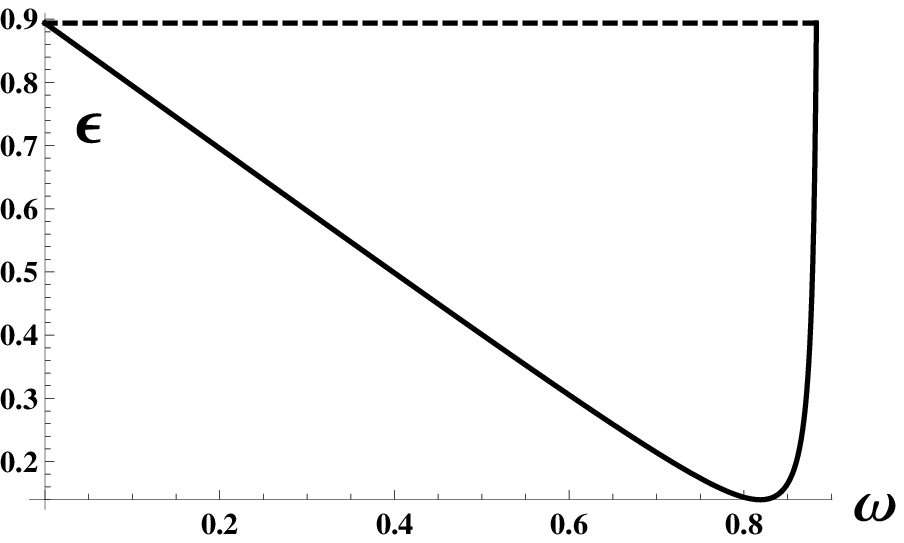}
%\end{center}

\caption {The regions of variation for the
radiative $\tau^-$ decay for different sets of kinematical
variables. The energies of the photon and pion $(\omega\,, \
\varepsilon)$ are given in GeV and the invariant variable\,
$t$-- in GeV\,$^2$. The line equations are defined by the
inequalities (6) and (7)\,.}
%\end{minipage}
\end{figure}
%\vspace{0.5cm}

The limits of variation of $c_1$ and $c_2$ are defined by the
condition of positiveness of the expression under the square root
in Eq.~(8) and they are shown in Fig.~3. A simple calculation
gives
$$c_{1-}\leq c_1\leq c_{1+}\,, \ \ -1\leq c_2\leq 1\,, \ \ c_{1\pm}=c_2c_{12}\pm s_2s_{12}\,,$$
where $s_{12}=\sin{\theta_{12}}.$ Besides, the integration in the
above limits gives
\begin{equation}\label{9}
\int\frac{dc_1}{K}= \int\frac{dc_2}{K}= \pi\,, \ \
\int\frac{c_1dc_1}{K}=\pi c_2c_{12}\,, \ \  \int\frac{c_2dc_2}{K}=
\pi c_1c_{12}\,.
\end{equation}
If $|M_{\gamma}|^2$ has no angular dependence, it corresponds to the
unpolarized case.

\vspace{0.4cm}

\begin{figure}
\captionstyle{flushleft}
\includegraphics[width=0.32\textwidth]{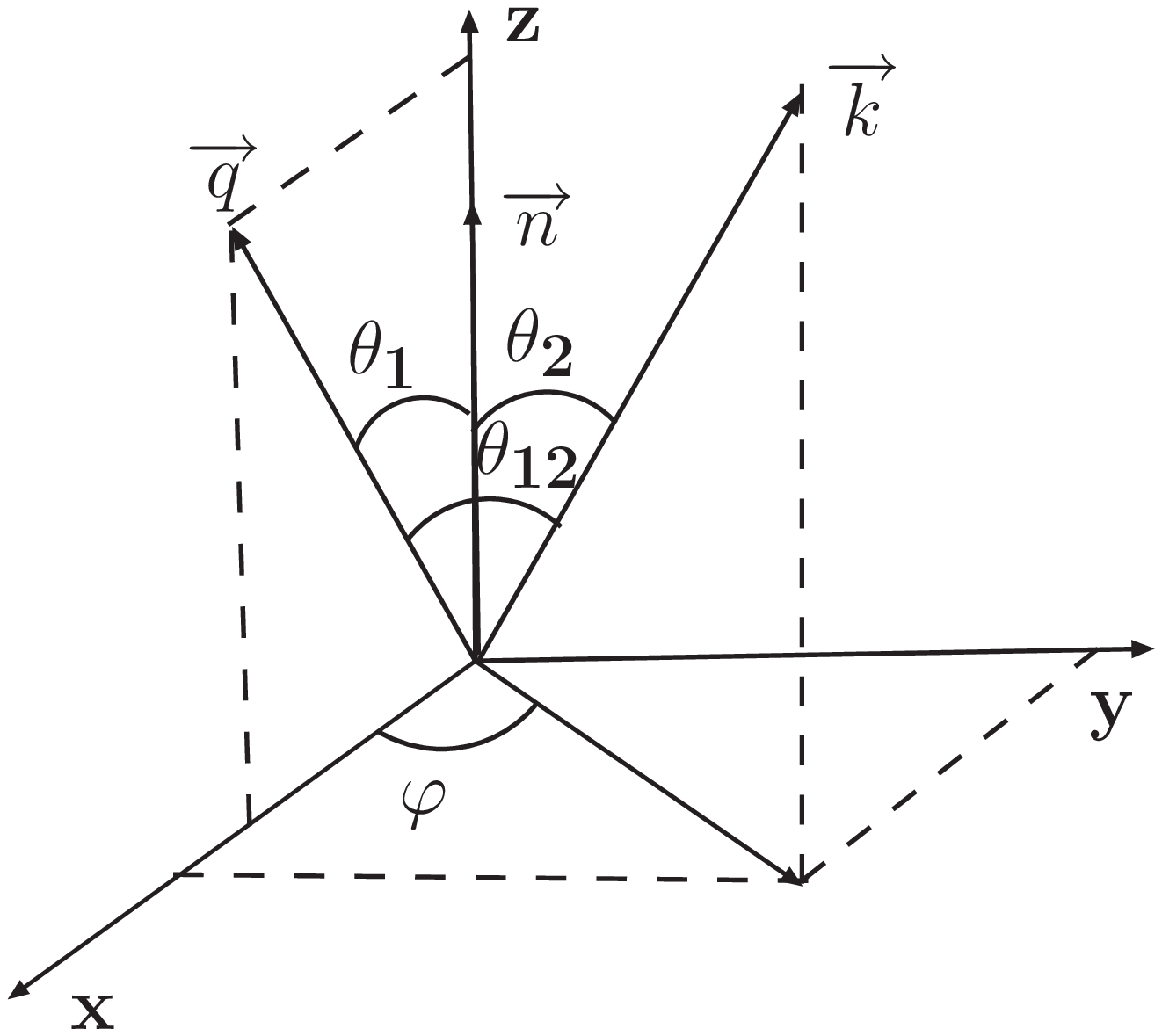}
%\hspace{0.4cm}
\includegraphics[width=0.32\textwidth]{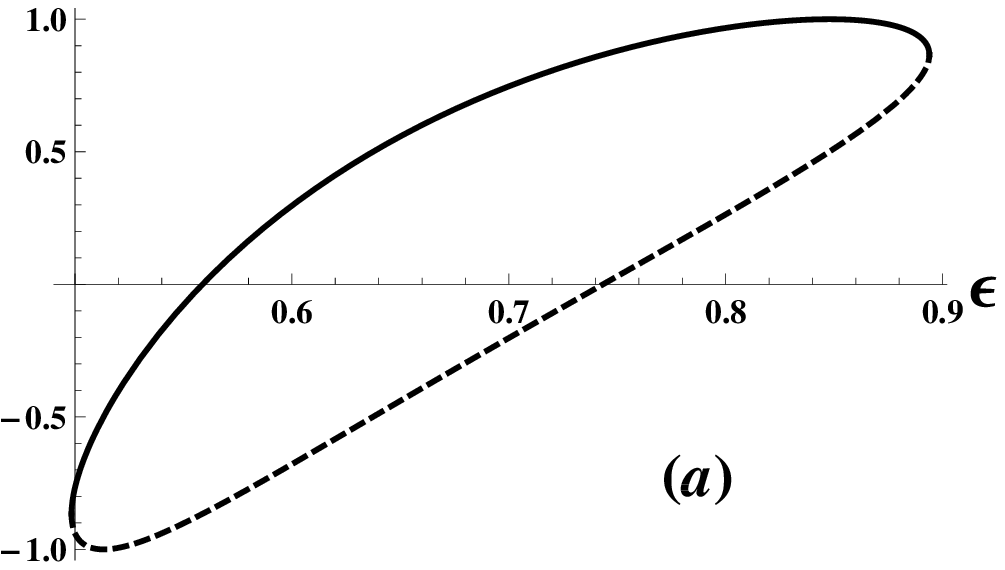}
\includegraphics[width=0.32\textwidth]{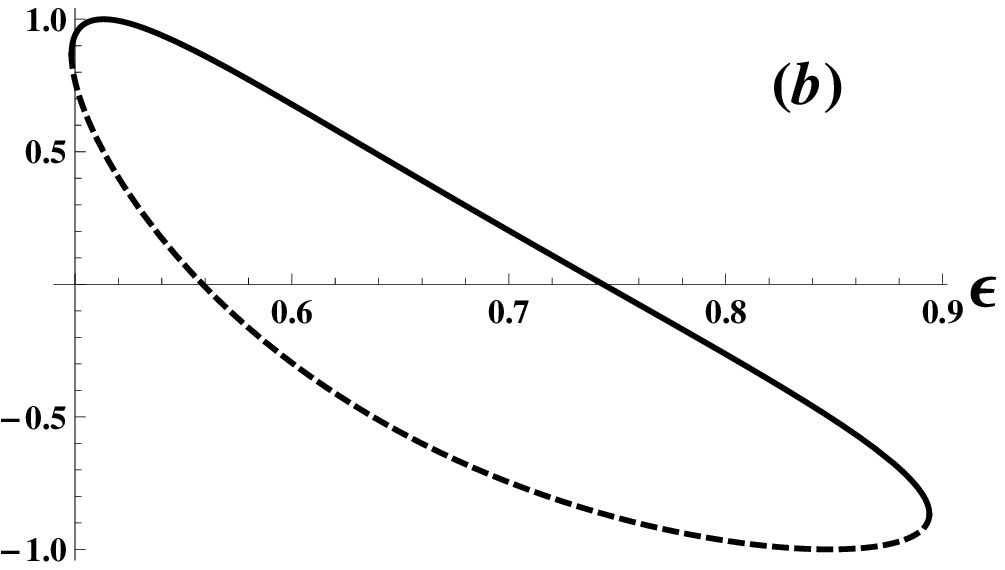}

\caption{Definition of the angles for the
case of the polarized radiative $\tau$ decay at rest (left
picture)\,; the limits of variation of $c_1$ at $\omega=0.4$ GeV
and $c_2=150^o$ (a) and $c_2=30^o$ (b): the solid (dashed) line
corresponds to $c_{1+}\, (c_{1-}).$}
\end{figure}

\vspace{0.5cm}

\subsection{Decay asymmetry and Stokes parameters}

To determine moduli and phases of the form factors it is necessary to measure some polarization observables.
The simplest of such observables are the so-called single-spin quantities: the asymmetries caused by the $\tau$
lepton polarization  and the photon Stokes parameters. We consider also the double-spin quantities: the dependence of the Stokes
parameters on  the $\tau$ lepton polarization.

Generally, the photon polarization properties  are described by its
Stokes parameters. At this point we have to clarify the
terminology used. The measurable Stokes parameters $\bar{\xi}_i \,,
i=1\,, 2\,, 3$ define the covariant spin-density matrix of a
photon in terms of quantities $\bar{\xi}_i$ and two independent
polarization 4-vectors $e_{1\mu}$ and $e_{2\mu}$ \cite{Berest}
\begin{equation}\label{10}
\rho_{\mu\nu}=\frac{1}{2}\big[e_{1\mu}e_{1\nu}+e_{2\mu}e_{2\nu}
+\bar{\xi}_1(e_{1\mu}e_{2\nu}+e_{1\nu}e_{2\mu})-i\bar{\xi}_2(e_{1\mu}e_{2\nu}-e_{1\nu}e_{2\mu})+\bar{\xi}_3(e_{1\mu}e_{1\nu}-e_{2\mu}e_{2\nu})
\big]\,,
\end{equation}
$$e^2_1=e^2_2=-1\,, \ \ (ke_1)=(ke_2)=0\,.$$

If the parameters $\bar{\xi}_i$ are measured, the matrix element
squared can be written as contraction of the current tensor
$T^{\mu\nu}$ and matrix $\rho_{\mu\nu},$
\begin{equation}\label{11}
|M_{\gamma}|^2 = T^{\mu\nu}\rho_{\mu\nu}\ ,
\end{equation}
where the current tensor $T^{\mu\nu}$ obeys the evident conditions
due to the electromagnetic current conservation
$$k_{\mu}T^{\mu\nu} = T^{\mu\nu}k_{\nu} =0 \ .$$

The polarization states of a real photon are described by two
independent pure space polarization vectors ${\bf{l}_1}$ and
${\bf{l}_2},$ which are both perpendicular to its 3-momentum ${\bf
k}$. In our case, in the rest system of the decaying $\tau^-$ lepton,
it is convenient to take ${\bf{l}_1}$ in the decay plane and
${\bf{l}_2}$ to be perpendicular to this plane. One can determine
two covariant polarization 4-vectors, which in the rest frame
coincide with ${\bf{l}_1}$ and ${\bf{l}_2}.$ These vectors are
defined as follows
\begin{equation}\label{12}
\varepsilon_1^{\mu}=e_1^{\mu}
-\frac{1}{N}\Big[(qp)-M^2\frac{(qk)}{(pk)}\Big]k^{\mu}\ , \
\varepsilon_2^{\mu}\equiv e_2^{\mu}=\frac{(\mu pqk)}{N}\ ,
\end{equation}
where $e_1^{\mu}$ and $N$ are defined after Eq.~(2).

It is easy to verify that in the rest frame
$$\varepsilon_{1\mu}=(0~, {\bf{l}_1})\,, \ \ \varepsilon_{2\mu}=(0~, {\bf{l}_2})\,, \
 \ {\bf{l}_1}^2={\bf{l}_2}^2=1\,,$$
$${\bf{l}_1}=\frac{{\bf q}-{\bf \hat k(\hat k q)}}{\sqrt{{\bf q^2}-{\bf (\hat k
q)}^2}}\,, \ \
 {\bf{l}_2}=[{\bf \hat k\, l_1}]\,, \ \
 {\bf \hat k}=\frac{{\bf k}}{\omega}\,, \
({\bf{l}_1}{\bf \hat k})=({\bf{l}_2}{\bf \hat k})=0\,.$$
Therefore, the set of the unit 3-vectors ${\bf l_1\,, ~ l_2}\,,$
and ${\bf \hat k}$ forms the right system of rectangular
coordinates. Due to the electromagnetic current conservation, we
can replace $ \varepsilon_{1\mu}$ by $e_{1\mu}$ in calculations of
any observable. Thus, in Eq.~(10) we will use polarization
4-vectors in form (12). Then the matrix element squared reads
\begin{equation}\label{13}
|M_{\gamma}|^2 = \frac{1}{2}\big[\Sigma+\Sigma_i\, \bar{\xi}_i\big]\,,
\end{equation}
where
$$\Sigma=T^{\mu\nu}(e_{1\mu}e_{1\nu}+e_{2\mu}e_{2\nu})\,, \ \ \Sigma_1=T^{\mu\nu}(e_{1\mu}e_{2\nu}+e_{1\nu}e_{2\mu})\,,$$
$$\Sigma_2=-i\,
T^{\mu\nu}(e_{1\mu}e_{2\nu}-e_{1\nu}e_{2\mu})\,, \ \
\Sigma_3=T^{\mu\nu}(e_{1\mu}e_{1\nu}-e_{2\mu}e_{2\nu})\,.$$ It is
obvious that parameters $\xi'_i$ depend on the properties of
detectors which analyze the polarization states of the photon and
do not depend on the production mechanism. On the contrary, quantities
$\Sigma$ and $\Sigma_i$ are defined only by the decay amplitude
and thus they determine  polarization properties of the photon
itself in the decay (1) \cite{Berest}. To study predictions of
different theoretical models for these quantities we can
write
$$|M_{\gamma}|^2 =\Sigma+\Sigma_i$$
instead of the expression (13).

For a polarized $\tau$ lepton the current tensor reads
$$T_{\mu\nu}=T^{^0}_{\mu\nu}+T^{^S}_{\mu\nu}\,,$$
where $T^{^S}_{\mu\nu}$ depends on the polarization 4-vector of
$\tau$ lepton and $T^{^0}_{\mu\nu}$ does not. In this case we can
write
$$\Sigma=\Sigma^{^0}+\Sigma^{^S}\,, \ \ \Sigma_i=\Sigma^{^0}_i+\Sigma^{^S}_i\,,$$
and define the physical quantities
\begin{equation}\label{14}
A^{^S}=\frac{\Sigma^{^S}d\,\Phi}{\Sigma^{^0}d\,\Phi}\,, \ \
\xi_i=\frac{\Sigma^{^0}_id\,\Phi}{\Sigma^{^0}d\,\Phi}\,, \ \
\xi^{^S}_i=\frac{\Sigma^{^S}_id\,\Phi}{\Sigma^{^0}d\,\Phi}\,,
\end{equation}
which completely describe the polarization effects in considered
decay.

The quantity $A^{^S}$ represents the polarization asymmetry of the
differential decay width caused by the polarization of $\tau$
lepton. The quantities $\xi_i$ are the
Stokes parameters of the photon itself provided that $\tau$ lepton
is unpolarized and the quantities $\xi^{^S}_i$ are the correlation
parameters describing influence of $\tau$ lepton polarization on
the photon Stokes parameters.

Thus, to analyze the polarization phenomena in the process (1) we
have to study both the spin-independent and spin-dependent parts
of the differential width and, in accordance with Eq.~(4), they
are
$$\frac{d\,\Gamma_0}{d\,\Phi}=g\Sigma^{^0}, \ \ \frac{d\,\Gamma^{^S}_0}{d\,\Phi}=g\Sigma^{^S},
\ \frac{d\,\Gamma_i}{d\,\Phi}=g\Sigma^{^0}_i, \
\frac{d\,\Gamma^{^S}_i}{d\,\Phi}=g\Sigma^{^S}_i, \
g=\frac{1}{4\,M(2\,\pi)^5}\,.$$ Note that, by partial integration
in numerators and denominator in the relations (14), we can define
and study also the corresponding reduced polarization parameters.

If we record the photon and pion energies, the parameters
$\xi_1(\omega,\epsilon)$ and $\xi_3(\omega,\epsilon)$ describe
linear polarizations of the photon and the parameter
$\xi_2(\omega,\epsilon)$ - the circular one. The last parameter
does not depend on the choice of two polarization vectors which,
in our work, are defined by the relations (12). On the contrary, each of
the parameters $\xi_1(\omega,\epsilon)$ and
$\xi_3(\omega,\epsilon)$ depends on the axes relative to which it
is defined, and only the quantity
$\sqrt{\xi_1(\omega,\epsilon)^2+\xi_3(\omega,\epsilon)^2}$ remains
invariant. In principle, one can choose the polarization vectors
$e\,'_1$ and $e\,'_2$ so that to vanish, for example,
$\xi\,'_1(\omega,\epsilon);$ then
$\xi\,'_3(\omega,\epsilon)=\sqrt{\xi_1(\omega,\epsilon)^2+\xi_3(\omega,\epsilon)^2}$
(and vice versa).

This statement can be verified by a simple rotation of the 4-vectors
$e_1$ and $e_2$ in the plane perpendicular to the direction ${\bf
k}$ \cite{Berest}
\begin{equation}\label{15}
e\,'_{1\mu}=e_{1\mu}\cos{\beta}-e_{2\mu}\sin{\beta}\,, \ \
e\,'_{2\mu}=e_{1\mu}\sin{\beta}+e_{2\mu}\cos{\beta}\,
\end{equation}
so that in terms of $e\,'_1$ and $e\,'_2$ we have, using
definitions (13) and (14)

$$\xi\,'_1(\omega,\epsilon)=\xi_1(\omega,\epsilon)\cos{2\beta}+\xi_3(\omega,\epsilon)\sin{2\beta}\,,
$$ $$
\xi\,'_3(\omega,\epsilon)=-\xi_1(\omega,\epsilon)\sin{2\beta}+\xi_3(\omega,\epsilon)\cos{2\beta}\,,
\ \ \xi\,'_2(\omega,\epsilon)=\xi_2(\omega,\epsilon)\,.$$ Taking,
for example,
$$\cos{2\beta}=\frac{\xi_1(\omega,\epsilon)}{\sqrt{\xi_1(\omega,\epsilon)^2+\xi_3(\omega,\epsilon)^2}}\,, \ \
\sin{2\beta}=\frac{\xi_3(\omega,\epsilon)}{\sqrt{\xi_1(\omega,\epsilon)^2+\xi_3(\omega,\epsilon)^2}}\,,$$
one easy obtains
$$ \xi\,'_1(\omega,\epsilon)=\sqrt{\xi_1(\omega,\epsilon)^2+\xi_3(\omega,\epsilon)^2}\,, \ \ \xi\,'_3(\omega,\epsilon)=0.$$

If the experimental setup allows to fix the decay plane, the
Stokes parameter $\xi_3$ defines the probability of the photon linear polarization
along two orthogonal directions: along ${\bf l_1}$ and
${\bf l_2}\,.$  If $\xi_3=1, (\xi_3=-1)$ the photon is fully
polarized along the direction ${\bf l_1}\,, ({\bf l_2})$ and its
polarization vector lies in the decay plane (perpendicular to it).
In general, the probability of the linear polarization in the
decay plane is $(1+\xi_3)/2$ and in the plane perpendicular to it
is $(1-\xi_3)/2.$

The parameter $\xi_1$ defines the probability of the linear polarization
in
the planes rotated by the angle $\phi =\pm 45~^o$ around the ${\bf k}$
-direction relative to the decay one. The full linear polarization
at $\phi=+45^o \ (-45^o)$ occurs at $\xi_1=1 \ (-1).$ The
corresponding probabilities, in a general case, are $(1+\xi_1)/2$
and $(1-\xi_1)/2,$ respectively.  Thus, we can say that the
circular polarization degree of a photon equals
$\xi_2(\omega,\epsilon)$ ( $\xi_2(\omega,\epsilon)=1~~(-1) $
corresponds to full right (left) circular polarization) and linear
one equals
$\sqrt{\xi_1(\omega,\epsilon)^2+\xi_3(\omega,\epsilon)^2}.$
Nevertheless, both parameters $\xi_1(\omega,\epsilon)$ and
$\xi_3(\omega,\epsilon)$ are measurable and carry different
information about the decay mechanism. But to define them
separately one has to determine the plane $({\bf q\,, k})$ in
every event. The same regards also reduced photon polarization
parameters, for example, $\xi_i(\omega)$ and so on.

\subsection{Polarization of $\tau$ lepton}

Before proceeding further, let us discuss briefly the possible
polarization states of a $\tau$ lepton created at $\tau$-factories
in electron-positron annihilation process with a longitudinally
polarized electron beam

$$e\,^- +e^+\rightarrow \tau\,^- +\tau^+.$$

Simple calculations in the lowest approximation of QED show
that polarization of $\tau$ arises if at least one of the
colliding beams is polarized. For example, in the case of a
longitudinally polarized electron beam, the $\tau^-$ lepton has
longitudinal and transverse polarizations in the reaction plane
relative to the $\tau$ lepton momentum direction (for the details
see Appendix B)
% this formulas requires the corrections!
\begin{equation}\label{16}
P^L=\frac{2\lambda\cos\theta}{Q}\ , \ \ P^T=\frac{4\lambda
M\sin\theta}{\sqrt{s}Q}\ , \
Q=1+\cos^2\theta+\frac{4M^2}{s}\sin^2\theta\ ,
\end{equation}
where $\lambda$ is the beam polarization degree , $\theta$ is the
angle between 3-momenta of the electron and $\tau^-$- lepton in
c.m.s. and $s$ is the total energy squared in c.m.s. If we select
and analyze the events with a longitudinally (transversally)
polarized $\tau^-$, then 3-vector ${\bf n}$ in the rest system
(see Fig.~3) lies in the annihilation reaction plane and it is
directed along (perpendicular to) the 3-momentum of $\tau^-$ in
c.m.s.

\section{Calculation of the current tensor $T^{\mu\nu}$}

The current tensor $T^{\mu\nu}$ contains three contributions --
the IB, the resonance ones and the interference between the IB and
resonance amplitudes. We divide every contribution into the
symmetric and antisymmetric parts relative to the Lorentz indices.
The symmetric part contributes to $\Sigma$, $\Sigma_1$ and
$\Sigma_3$ whereas the antisymmetric one -- to $\Sigma_2$ only.

Below, in the formulas for the different parts of the current
tensor we omitted the terms proportional to the 4-vectors
$k_{\mu}$ and $k_{\nu}$ since they do not contribute to the
observables.

\subsection{Inner bremsstrahlung contribution}

For the IB contribution we have
$$
T_{IB}^{\mu\nu}=4Z^2M^2\big[S_{IB}^{\mu\nu}+iA_{IB}^{\mu\nu}\big]\
,
$$ where the symmetric part reads
\begin{equation}\label{17}
S_{IB}^{\mu\nu}=\frac{(qk)-(pk)}{(pk)^2}\big[(pk)+M(kS)\big]g^{\mu\nu}
+\frac{N^2}{(pk)^2(qk)^2}\big[M^2-m^2+2M(p'S)\big]e_1^{\mu}e_1^{\nu}+
\end{equation}
$$\frac{NM}{(pk)^2(qk)}\big[(e_1l_p)^{\mu\nu}-(e_1l_q)^{\mu\nu}\big] \ ,$$
and antisymmetric one is
\begin{equation}\label{18}
A_{IB}^{\mu\nu}=\frac{(pk)-(qk)}{(pk)^2}[M(\mu\nu kS)-(\mu\nu
pk)]+\frac{N^2}{(pk)^2(qk)}[e_1e_2]^{\mu\nu}-
\end{equation}
$$\frac{MN}{(pk)^2(qk)}[e_1^{\mu}(\nu p'kS)- e_1^{\nu}(\mu p'kS)]\ .
$$
Here we used the following notation
$$ l_p^{\mu}=(pk)S^{\mu}-(kS)p^{\mu}\ , \ \ l_q^{\mu}=(qk)S^{\mu}-(kS)q^{\mu}\ , $$
$$ (ab)^{\mu\nu}=a^{\mu}b^{\nu}+a^{\nu}b^{\mu}\ , \ \ [ab]^{\mu\nu}=a^{\mu}b^{\nu}-a^{\nu}b^{\mu}\,, \ \ (kl_p)=(kl_q)=0\,.$$
Note that the form of the antisymmetric part can be written in different equivalent forms. Indeed, one can
derive another form using the well known relation
$$
g^{\alpha\beta}(\mu\nu\lambda\rho)=g^{\alpha\mu}(\beta\nu\lambda\rho)+g^{\alpha\nu}(\mu\beta\lambda\rho)
+g^{\alpha\lambda}(\mu\nu\beta\rho)+g^{\alpha\rho}(\mu\nu\lambda\beta)\
. $$

\subsection{Resonance contribution}

As concern the resonance contribution into the current tensor
$T^{\mu\nu},$ we write it in the form
$$T_R^{\mu\nu}=\frac{8Z^2}{M^4}\big[S_R^{\mu\nu}+iA_R^{\mu\nu}\big]\ ,$$
where both the symmetric and antisymmetric parts include four
independent pieces. They are proportional to $|a(t)|^2, \
|v(t)|^2, \ Re(a(t)v^*(t))$ and $ Im(a(t)v^*(t)).$ Denoting the
respective symmetrical pieces as $ S_{Ra},\ S_{Rv},\ S_{Rr}$ and $
S_{Ri}$ and the antisymmetrical ones as $ A_{Ra},\ A_{Rv},\
A_{Rr}$ and $ A_{Ri},$ we have
\begin{equation}\label{19}
S_{Ra}^{\mu\nu}=|a(t)|^2\big\{(qk)^2[M(p'S)-(pp')\big]g^{\mu\nu}+2N^2e_1^{\mu}e_1^{\nu}+NM(e_1l_q)^{\mu\nu}\big\}\
,
\end{equation}
\begin{equation}\label{20}
A_{Ra}^{\mu\nu}=|a(t)|^2(qk)\big\{(qp')(\mu\nu pk)-(qp)(\mu\nu
p'k)-M[(qp')(\mu\nu Sk)-(qS)(\mu\nu p'k)]\big\}\ .
\end{equation}
\begin{equation}\label{21}
S_{Rv}^{\mu\nu}=|v(t)|^2\big\{\big[M(p'S)-(pp')\big](qk)^2g_{\mu\nu}+2N^2e_2^{\mu}e_2^{\nu}-NM\big[
e_2^{\mu}(\nu qkS)+e_2^{\nu}(\mu qkS)\big]\big\}\ ,
\end{equation}
\begin{equation}\label{22}
A_{Rv}^{\mu\nu}=|v(t)|^2(\mu\nu
qk)\big\{(qk)[(pq)-(pk)]-m^2(pk)-M[(qp')(kS)-(p'k)(qS)]\}\ ,
\end{equation}
\begin{equation}\label{23}
S_{Rr}^{\mu\nu}=2Re\{a^*(t)v(t)\}g^{\mu\nu}(qk)\big\{(qk)\big[(pk)-(pq)\big]+m^2(pk)
\end{equation}
$$-M\big[(kp')(Sq)-(Sk)(p'q)\big]\big\}\ ,$$
\begin{equation}\label{24}
A_{Rr}^{\mu\nu}=2Re\{a^*(t)v(t)\}\big\{(qk)[(pp')-M(p'S)](\mu\nu
qk)-N^2[e_1e_2]^{\mu\nu}+
\end{equation}
$$\frac{1}{2}NM\big(e_1^{\mu}(\nu qkS)-e_1^{\nu}(\mu qkS)+[e_2l_q]^{\mu\nu}\big)\big\}\
,$$
\begin{equation}\label{25}
S_{Ri}^{\mu\nu}=2Im\{a^*(t)v(t)\}\big\{\frac{1}{2}NM\big[-(e_2l_q)^{\mu\nu}
+e_1^{\mu}(\nu qkS)+e_1^{\nu}(\mu
qkS)\big]-N^2(e_1e_2)^{\mu\nu}\big\}\ ,
\end{equation}
\begin{equation}\label{26}
A_{Ri}^{\mu\nu}=0\ .
\end{equation}

The interference between the vector and axial-vector contributions
of the resonance amplitudes is sensitive to the relative sign of
the axial-vector and vector couplings and its separation can be
used to fix the sign of the ratio $f_V(0)/f_A(0).$

\subsection{IB -- Resonance interference}

The interference between the IB and resonance amplitudes is more
sensitive to all the resonance parameters because $a(t)$ and
$v(t)$ enter it linearly. It is very important to find such a
polarization observable where the interference contribution would
be enhanced relative to the background created by pure IB and
resonance contributions. For the current tensor caused by the IB -
resonance interference we define
$$T_{IR}^{\mu\nu}=\frac{8Z^2}{M(pk)}\big[S_{IR}^{\mu\nu}+iA_{IR}^{\mu\nu}\big]\ .$$
Again we have four symmetric and antisymmetric terms, which read
\begin{equation}\label{27}
S_{IRra}^{\mu\nu}=Re(a(t))\Big\{(qk)[(pk)(Sp')-(kS)(pp')-M(p'k)]g^{\mu\nu}+\frac{2N^2}{(qk)}[M+(Sp')]e_1^{\mu}e_1^{\nu}+
\end{equation}
$$\frac{N(pp')}{(qk)}(e_1l_q)^{\mu\nu}+N(e_1l_p)^{\mu\nu}\Big\}\
,$$
\begin{equation}\label{28}
A_{IRra}^{\mu\nu}=Re(a(t))\Big\{\big[(qS)\big(2(pk)-(qk)+(pq)\big)-M(qk)\big](\mu\nu
pk)+
\end{equation}
$$\big[(pq)\big(2(pk)-(qk)+(pq)\big)-M^2(m^2+(qk))\big](\mu\nu kS)+M\big[(qk)-M(qS)\big](\mu\nu qk)\Big\}\ ,$$

\begin{equation}\label{29}
S_{IRia}^{\mu\nu}=-Im(a(t))N\Big[(qk)(e_2S)g^{\mu\nu}+\frac{(e_1Q)^{\mu\nu}}{(qk)}\Big]\
,
\end{equation}
\begin{equation}\label{30}
A_{IRia}^{\mu\nu}=Im(a(t))N\Big[[e_1l_p]^{\mu\nu}-\bigg(1+\frac{(pp')}{(qk)}\bigg)[e_1l_q]^{\mu\nu}\Big]\
,
\end{equation}
 where
$$Q^{\mu}=N(e_2S)q^{\mu}-(qk)(Spq\mu)\ ,$$.
\begin{equation}\label{31}
S_{IRrv}^{\mu\nu}=Re(v(t))\Big\{(p'k)\big[M(qk)+(pq)(kS)-(pk)(qS)\big]g^{\mu\nu}+
2\frac{N^2}{(qk)}(qS)e_1^{\mu}e_1^{\nu}+
\end{equation}
$$N\Big[\frac{(pq)}{(qk)}\big(e_1l_q\big)^{\mu\nu}-\Big(\frac{1}{2}+\frac{m^2}{(qk)}\Big)
\big(e_1l_p\big)^{\mu\nu}+\frac{1}{2}\big(e_2^{\mu}(\nu
pkS)+e_2^{\nu}(\mu pkS)\big)\Big]\Big\}\ , $$

\begin{equation}\label{32}
A_{IRrv}^{\mu\nu}=Re(v(t))\Big\{\big[M(p'k)+(pp')(kS)-(pk)(p'S)\big](\mu\nu
qk)-\frac{N^2}{(qk)}[M+(p'S)][e_1e_2]^{\mu\nu}+
\end{equation}
$$N\Big[\frac{(pp')}{(qk)}\big[e_2l_q\big]^{\mu\nu}
+\big[e_2l_p\big]^{\mu\nu} \Big]\Big\}\ , $$

\begin{equation}\label{33}
S_{IRiv}^{\mu\nu}=Im(v(t))N\Big\{-\frac{N}{(qk)}[M+(p'S)]\big(e_1e_2\big)^{\mu\nu}+
\Big(\frac{1}{2}+\frac{(pp')}{(qk)}\Big)\big(e_1^{\mu}(\nu
qkS)+e_1^{\nu}(\mu qkS)\big)-
\end{equation}
$$\frac{1}{2}\big(e_2(2l_p-l_q)\big)^{\mu\nu}\Big\}\ ,$$

\begin{equation}\label{34}
A_{IRiv}^{\mu\nu}=Im(v(t))N\Big\{\Big(\frac{1}{2}+\frac{m^2}{(qk)}\Big)\big[e_1l_p\big]^{\mu\nu}
-\frac{(pq)}{(qk)}\big[e_1l_q\big]^{\mu\nu}+
\end{equation}
$$\frac{1}{2}\big[e_2^{\mu}(\nu pkS)-e_2^{\nu}(\mu pkS)\big]
-\big[e_2^{\mu}(\nu qkS)-e_2^{\nu}(\mu qkS)\big] \Big\}\ . $$

In the above formulas we omitted terms proportional to the 4-vectors $k_{\mu}$ and $k_{\nu}$
since they do not contribute to any observables.

The expression for the current tensor $T^{\mu\nu}$ allows us to
derive all the polarization observables in the Lorentz invariant form
by contracting this tensor with an appropriate
combination of 4-vectors $e_1$ and $e_2.$ The set of needed
formulas reads
$$ e_1^2=e_2^2=-1\ , \ \ (e_1e_2)=0\ , $$
$$(e_1l_p)=-(e_2pkS)=\frac{1}{N}\Big\{(Sq)(pk)^2+(Sk)\big[M^2(qk)-(pq)(pk)\big]\Big\}\ ,$$
$$(e_2l_p)=(e_1pkS)=\frac{(pk)}{N}(Spqk)\ ,$$
\begin{equation}\label{35}
(e_1l_q)=-(e_2qkS)=\frac{1}{N}\Big\{(Sq)(pk)(qk)+(Sk)\big[(pq)(qk)-m^2(pk)\big]\Big\}\,,
\end{equation}
$$(e_2l_q)=(e_1qkS)=\frac{(qk)}{N}(Spqk)\ ,$$
$$(e_1e_2qk)=-(qk)\ , \ (e_1e_2pk)=-(pk)\ , \ (e_1e_2Sk)=-(Sk)\ ,$$
$$ (e_1Q)=\frac{[m^2(pk)-(pq)(qk)]}{N}(Spqk)\ ,$$
$$ (e_2Q)=\frac{(qk)}{N}\Big\{(kS)\big[(pq)^2-M^2m^2\big]+(qS)\big[M^2(qk)-(pq)(pk)\big]\Big\}\ .$$

In the presence of a polarized $\tau$ lepton the structure of the
differential width and the Stokes parameters of the photon are
much richer. As we saw below, in the polarized
electron-positron annihilation the created $\tau$ leptons have
essential longitudinal or transverse polarizations. In both cases,
in the rest system of $\tau,$ its polarization 4-vector
$S^{\mu}=(0,{\bf n}),$ and choosing a coordinate system as shown in
Fig.~3 one has
\begin{equation}\label{36}
(qS)=-|{\bf q}|c_1\ , \ \ (kS)=-\omega c_2\ ,
\end{equation}
$$ (Spqk)=M\big({\bf n\,[\,q\,k\,]}\big)= sign(\sin\varphi)M|{\bf q}|\omega\sqrt{1-c_1^2-c_2^2-c_{12}^2+2c_1c_2c_{12}}\ ,$$
where $\sin{\varphi}$ defines the $y$-component of the 3-vector
$|{\bf k}|,$ namely $k_y=\omega\sin{\theta_2}\sin{\varphi}\ .$

If one sums events with all possible values of the azimuthal angle
$\varphi,$ the spin correlation $(Spqk),$ which is perpendicular
to the plane $({\bf q\,,k\,}),$ does not contribute.  On the other
hand, spin correlations in the plane $({\bf q\,,k\,}),$ caused by the
$(Sq)$ and $(Sk)$ terms, being integrated over $c_1$ (or over
$c_2$), are always proportional to $c_2$ (or $c_1$) as it follows
from the relations (9) and (36).

\section{Axial-vector and vector form factors in the R$\chi$T}

From the Lagrangians in Appendix~A one can obtain the photon
interaction with the pseudoscalar mesons $P$ in the following form
\begin{eqnarray}
{\cal L}_{\gamma P P } \, & = & \,   i e  B^\mu \, {\rm Tr}\,
 \Big( Q, \, [\Phi, \, \partial_\mu \Phi ]   \Big) = ieB^\mu (\pi^+ \stackrel{\leftrightarrow}{\partial_\mu} \pi^- + K^+ \stackrel{\leftrightarrow}{\partial_\mu}
 K^-),
 \label{eq:pi-pi-gamma}
 \\
\stackrel{\leftrightarrow}{\partial_\mu} \,  & \equiv &   \,
\stackrel{\rightarrow}{\partial_\mu}\, -
\,\stackrel{\leftarrow}{\partial_\mu}\,. \nonumber
\end{eqnarray}

The axial transition $W^\pm \to \pi^\pm  \,  (K^\pm)$ is described
by
\begin{eqnarray}
{\cal L}_{W P } &=& -   \frac{g F}{2} {\rm Tr}\,
 \Big( W_\mu  \partial^\mu \Phi \Big)
\nonumber \\
&=&   \frac{g F}{2} \,   W_\mu^+ \, (V_{ud} \partial^\mu \pi^- +
V_{us}
\partial^\mu K^-) \, + \, {\rm h.c.}
\label{eq:W-pi}
\end{eqnarray}

The vertex with an additional photon, $W^\pm \to \gamma \pi^\pm \,
(K^\pm)$, is generated from
 \begin{eqnarray}
{\cal L}_{W P \gamma } &=& -   \frac{i}{2} eg F \, B^\mu \, {\rm
Tr}\,
 \Big( [Q, \, \Phi] W_\mu \Big)
\nonumber \\
&=&   \frac{i}{2}eg F \,  B^\mu \,    W_\mu^+ \, (V_{ud} \pi^- +
V_{us} K^-) \, + \, {\rm h.c.} \label{eq:W-pi-gamma}
\end{eqnarray}

To evaluate the resonance contribution to the axial-vector form
factor one needs the vertex of the  $W^\pm \to  a_1^\pm \,
(K_1^\pm)$ transition
\begin{eqnarray}
{\cal L}_{W A } &=&    -  \frac{1}{4} g F_A  \,  {\rm Tr}\,
(W_{\mu \nu} A^{\mu \nu})
 \nonumber \\
&=& -   \frac{gF_A}{2} \,   \partial_\mu W_\nu^+ \, (V_{ud} a_1^{-
\, \mu \nu } + V_{us} K_1^{- \, \mu \nu } )\, + \, {\rm h.c.} \,.
\label{eq:W-a_1}
\end{eqnarray}

The transition $a_1^\pm \, (K_1^\pm) \to \gamma \pi^\pm \,
(K^\pm)$ is described by
\begin{equation}
{\cal L}_{ A  P \gamma} =     -  i \frac{e F_A}{2 F} \, F^{\mu \nu
} \, ( \pi^- a_{1 \, \mu \nu}^+ + K^- K_{1 \, \mu \nu}^+  ) \, +
\, {\rm h.c.} \label{eq:a_1-pi-gamma}
\end{equation}

The transition $W^\pm \to \pi^\pm \rho^0$ is generated from the
Lagrangian
 \begin{equation}
{\cal L}_{W P V }=    -  i \frac{g G_V}{\sqrt{2} F} \,  {\rm Tr}\,
\Big( V^{\mu \nu}\,  [W_\mu, \,  \partial_\nu \Phi]  \Big) +
 i \frac{g F_V}{4 \sqrt{2} F} \,  {\rm Tr}\,  \Big(  V^{\mu \nu}\,  [W_{\mu  \nu }\, , \Phi]   \Big)
\label{eq:W-V-Phi}
\end{equation}
with the notation
\begin{equation}
W_{\mu \nu}\,  =  \,   W_{\mu \nu}^+ T_+ +   W_{\mu \nu}^- T_-\,,
\qquad \qquad   W_{\mu \nu}^\pm \, \equiv  \,  \partial_\mu
W_\nu^\pm -
\partial_\nu W_\mu^\pm\,.
\label{eq:W_notation}
 \end{equation}

Keeping in Eq.~(\ref{eq:W-V-Phi}) the contribution from the
neutral vector mesons, one obtains
\begin{eqnarray}
{\cal L}_{W P V  } &=&       i \frac{g G_V}{\sqrt{2} F} \, W_\mu^+
[-\sqrt{2} V_{ud} \partial_\nu \pi^-  \rho^{0 \, \mu \nu}  +
V_{us} \partial_\nu K^-   (\phi - \frac{1}{\sqrt{2}} \rho^0 -  \frac{1}{\sqrt{2}} \omega)^{\mu \nu }  ]   \label{eq:W-rho-pi}   \\
&& - i  \frac{g F_V}{4 \sqrt{2} F} \, W_{\mu \nu}^+ [-\sqrt{2}
V_{ud} \pi^-  \rho^{0 \, \mu \nu}  + V_{us}  K^-   (\phi -
\frac{1}{\sqrt{2}} \rho^0 -  \frac{1}{\sqrt{2}} \omega)^{\mu \nu }
]  + {\rm h. \ c.} \nonumber
 \end{eqnarray}

Finally, one needs the term describing a transition of the neutral
vector mesons to photon
\begin{equation}
{\cal L}_{V \gamma } =    - i \frac{e F_V}{\sqrt{2} F} \,  F^{\mu
\nu} {\rm Tr}\,  \Big( V_{\mu \nu}\,  Q  \Big)  = e  F_V F^{\mu
\nu}  \Big( \frac{1}{2} \rho^0_{\mu \nu} +  \frac{1}{6}
\omega_{\mu \nu} - \frac{1}{3 \sqrt{2}} \phi_{\mu \nu} \Big)\,
\label{eq:rho-gamma}
 \end{equation}
with $F^{\mu \nu} = \partial^\mu B^\nu - \partial^\nu B^\mu$.

Collecting vertices from (\ref{eq:W-a_1}), (\ref{eq:a_1-pi-gamma})
for the transition $W^- \to a_1^- \to \pi^- \gamma$, and vertices
from (\ref{eq:W-rho-pi}), (\ref{eq:rho-gamma}) for the transition
$W^- \to \pi^- \rho^0 \to \pi^- \gamma$, we obtain the
axial-vector form factor in the form
\begin{equation}
f_A (t) = \frac{\sqrt{2}  m_{\pi \pm} }{F_\pi} \biggl[
\frac{F_A^2}{m_a^2 - t - i m_a \Gamma_a (t)} + \frac{F_V (2 G_V -
F_V)}{m_\rho^2} \biggr] \, , \label{eq:axial FF}
\end{equation}
where $\Gamma_a (t)$ is the decay width of the $a_1$-meson.

Note that the axial-vector form factor in Eq.~(\ref{eq:axial FF})
is normalized at $t=0$ as
\begin{equation}
f_A (0) = \frac{\sqrt{2}  m_{\pi \pm} }{F_\pi} \biggl[
\frac{F_A^2}{m_a^2} + \frac{F_V (2 G_V - F_V)}{m_\rho^2} \biggr]
\,, \label{eq:axial FF_0}
\end{equation}
which is consistent with the chiral expansion in order ${\cal
O}(p^4)$ (see also Refs.~\cite{Geng_2004, Unterdorfer_2008}) in
terms of the low-energy constants,
\begin{eqnarray}
f_A (0) & = &  \frac{4 \sqrt{2} m_{\pi^\pm} }{F_\pi} ( L_9 + L_{10}) ,  \nonumber \\
L_9  & = & \frac{F_V G_V}{2 m_\rho^2},  \qquad \qquad  \qquad
L_{10 } = \frac{F_A^2}{4 m_a^2} -  \frac{F_V^2 }{4 m_\rho^2}\,.
\label{eq:LECs}
\end{eqnarray}
The expressions for $L_9, \ L_{10}$ follow from the resonance
saturation of the low-energy constants \cite{EckerNP321,
EckerPL223}.

The masses of the $\rho (770)$ and $a_1 (1260)$ mesons in
Eq.~(\ref{eq:axial FF}) are \  $m_\rho= 0.7755 $ GeV and $m_a =
1.230$ GeV. The width of the $a_1 (1260)$ meson in
Eq.~(\ref{eq:axial FF}) is taken from Ref.~\cite{Ecker_2003}:
\begin{eqnarray}
&& \Gamma_a (t) = \Gamma_0 \ {g(t)}/{g(m_a^2)},  \\
&& g(t) = \bigl(1.623 \ t +10.38 -\frac{9.23}{t} +
\frac{0.65}{t^2}\bigr) \ \theta(t-(m_\rho +m_\pi)^2)
\nonumber \\
&&+   4.1 (t-9m_\pi^2)^3 \bigl[1.0 -3.3 (t-9m_\pi^2) +5.8
(t-9m_\pi^2)^2 \bigr] \  \theta (t-9m_\pi^2) \theta ((m_\rho
+m_\pi)^2 -t) \, . \nonumber
\end{eqnarray}
It is implied in this equation that $t$  is in GeV$^2$, masses are
in GeV ($m_\pi =m_{\pi \pm}$) and all numbers are in appropriate
powers of GeV. An alternative form of the off-mass-shell $a_1$ decay width is proposed in Ref.~\cite{Dumm:2009va}.

The values of the coupling constants $F_A, \, F_V $ and $G_V$ are
presented in Table \ref{tab:axial}.
The constants $F_V$ and $G_V$ are obtained from the experimental information~\cite{PDG} on
the $\rho \to e^+ e^-$ and $\rho \to \pi \pi$ decay widths:  $\Gamma({\rho \to e^+ e^-}) = 7.04 \pm 0.06$ keV and
$\Gamma({\rho  \to \pi^+ \pi^-}) = 146.2 \pm 0.7$ MeV.
To find $F_V, \, G_V$  one can use the tree-level relations
%\begin{equation}
$$\Gamma({\rho \to e^+ e^-}) = \frac{e^4 F_V^2}{12 \pi m_\rho} \, , \qquad  \qquad
\Gamma({\rho  \to \pi^+ \pi^-}) = \frac{G_V^2 m_\rho^3}{48 \pi F_\pi^4} \Big( 1 - \frac{4 m_\pi^2}{m_\rho^2} \Big)^{3/2} \,.$$
%\label{eq:decay_widths}
%\end{equation}
The constant $F_A$ is then calculated from Eq.~(\ref{eq:axial FF_0}) using the
average value $f_A (0)_{exp}=0.0119 \pm 0.0001$
measured in the radiative pion decays~\cite{PDG}. The constants $F_V, \, G_V $ and $F_A$, calculated for central values of the data, are called  hereafter ``set 1'' and are shown in Table \ref{tab:axial}.

As another option we choose theoretically motivated values of the constants from Ref.~\cite{EckerPL223}. In particular, the relations $F_V =2 G_V$ and $F_V G_V = F_\pi^2$ are suggested there. The corresponding parameters are called ``set 2'' and are also given in Table \ref{tab:axial}.

\begin{table}[tbh]
\onelinecaptionsfalse
 \captionstyle{flushleft}
\begin{center}
%\centering
\begin{tabular}{|c|c|c|c|}
\hline
   & $F_A$  &  $F_V$  & $G_V$    \\
\hline
 ~set 1 ~& ~0.1368 GeV~  & ~0.1564 GeV ~& ~0.06514 GeV ~   \\
%\hline
 set 2 & $F_\pi$ & $\sqrt{2} F_\pi $ & $F_\pi/\sqrt{2}$  \\
\hline
\end{tabular}
\caption{Two sets of the coupling constants. Values for set 1
are calculated from the $\rho \to e^+ e^-, \, \rho \to \pi^+ \pi^-$ decay widths and $f_A (0)_{exp}$ (see the text). Values for set 2 are chosen according to Ref.~\cite{EckerPL223}, and $F_\pi=0.0924$ GeV. }
\label{tab:axial}
\end{center}
\end{table}

%%%%%%%%%%%%%%%%%%%%%%%%%%%%%%%%%%%%%%%%%%%%%%%%%%%
%%%%%%%%%%%%%%%%%%%%%%%%%%%%%%%%%%%%%%%%%%%%%%%%%%
%%%%%%%%%%%%%%%%%%%%%%%%%%%%%%%%%%%%%%%%%%%%%%%%%%%

In the calculation of the vector form factor $f_V(t)$ one needs
the transition $W^\pm \to \rho^\pm \to \pi^\pm  \gamma$, which
involves an odd-intrinsic-parity (anomalous) vertex. For the latter we use the
vector (or Proca) representation for the spin-one fields. As shown
in Ref.~\cite{EckerPLB237} (see also Ref.~\cite{Cirigliano_2006}),
the use of the vector field $V^\mu$ instead of the antisymmetric
tensor field $V^{\mu \nu}$ in the description of the spin-one
resonances ensures the correct behavior of the Green functions to
order ${\cal O}(p^6)$, while the tensor formulation would require
additional local terms (see also the discussion in Appendix~F of
Ref.~\cite{Dubinsky_2005}).

Thus we choose the Lagrangian~\cite{EckerPLB237,Cirigliano_2006}
\begin{equation}
{\cal L}_{V P \gamma } =    - h_V \epsilon_{\mu \nu \alpha \beta}
\, {\rm Tr}\, (V^{\mu} \{ u^\nu, \, f_+^{\alpha \beta} \}) \, = \,
\frac{4 \sqrt{2} \, e\, h_V}{3F_\pi} \epsilon_{\mu \nu \alpha
\beta}
\partial^\alpha B^\beta \vec{\rho}^{\, \mu} \partial^\nu{\vec\pi}
\label{eq:rho-pi-gamma}
\end{equation}
with the coupling constant $h_V$.

For the $W^\pm \to \rho^\pm \, (K^{* \pm})$  vertex, in the vector
formulation, one has
\begin{eqnarray}
{\cal L}_{W V } &=&    -  \frac{1}{4} g  \frac{F_V}{m_\rho}  \,
{\rm Tr}\, (W_{\mu \nu} \partial^\mu V^\nu)
 \nonumber \\
&=& -   \frac{g F_V}{4 m_\rho}\,  W_{\mu \nu}^+   \, (V_{ud}
\partial^\mu \rho^{- \, \nu } + V_{us} \partial^\mu K^{*- \, \nu }) \,
+ \,  h.c. \label{eq:WV_vector}
\end{eqnarray}

Using (\ref{eq:rho-pi-gamma}), (\ref{eq:WV_vector}) and adding the
vertex (\ref{eq:WZW_2}), we obtain the vector form factor
\begin{equation}
f_V (t) =  \frac{ \sqrt{2} m_{\pi \pm} }{F_\pi} \biggl[
\frac{N_C}{24 \pi^2} + \frac{4 \sqrt{2} h_V F_V}{3 m_\rho}
 \frac{t}{m_\rho^2 -t -i m_\rho  \Gamma_\rho (t)} \biggr].
 \label{eq:vector FF}
\end{equation}

The width of the off-mass-shell $\rho$ meson can be calculated from the  interaction Lagrangian ${\cal L}_{int}^R$ in Eqs.~(\ref{eq:lagr-R-inter}).
It is written in the form
\begin{equation}
\Gamma_\rho (t) = \frac{G_V^2 m_\rho t}{48 \pi F_\pi^4 }\biggl[
\Big(1-\frac{4 m_\pi^2}{t} \Big)^{3/2} \theta(t-4 m_\pi^2) + \frac{1}{2}
\Big(1-\frac{4 m_K^2}{t} \Big)^{3/2} \theta(t-4 m_K^2)\biggr]\,,
\label{eq:rho_width}
\end{equation}
where $m_K=0.4937$ GeV is the mass of the $K^\pm$ meson. Other contributions to the width, coming, for example, from the four-pion decay of the $\rho$, are neglected in (\ref{eq:rho_width}).

The coupling constant $h_V$ can be fixed from the
decay width $\Gamma (\rho^\pm \to \pi^\pm \gamma) = 68 \pm 7$
keV~\cite{PDG}. Then from the equation
%\begin{equation}
$$\Gamma (\rho^\pm \to \pi^\pm \gamma) = \frac{e^2  m_\rho^3 h_V^2}{27 \pi F_\pi^2} \Big( 1 - \frac{m_\pi^2}{m_\rho^2} \Big)^3$$
%\label{eq:h_V}
%\end{equation}
we obtain $h_V = 0.036$.
Alternatively, $h_V$ can be constrained from the high-energy
behavior of the vector form factor. Such constraints have been
used in Refs.~{\cite{D93,G10}}. According to the asymptotic
predictions of the perturbative QCD~\cite{Lepage}, at $t \to -
\infty$ the form factor behaves as $f_V (t) \sim {\rm const} /t$.
Imposing this condition on $f_V (t)$ in Eq.~(\ref{eq:vector FF})
one obtains
\begin{equation}
h_V = \frac{N_c m_\rho}{32 \pi^2 \sqrt{2} F_V} \ .
\label{eq:h_V}
\end{equation}
This yields $h_V$ = 0.033 (0.040) for $F_V$ from the set 1 (set 2) in
Table \ref{tab:axial}. These values are close to $h_V = 0.036$
derived from the $\rho \to \pi \gamma$ decay width, and the latter value is used in our calculations. Recently an approach based on Lagrangian with bilinear in resonances terms has been applied in Ref.~\cite{Roig:2014}, and the high-energy constraints in the anomalous QCD sector have been formulated in the tensor representation for the spin-one fields. Note that in this approach the vector form factor depends on several parameters which can be determined from these high-energy constraints.

\section{Calculation of differential decay width, Stokes parameters, polarization asymmetry and spin-correlation
parameters}
% on the squared invariant mass of $\pi\, \gamma$ system}

\subsection{The t-dependence in the case of unpolarized $\tau^-$}

Because the vector and axial-vector  form factors depend on
invariant mass of the $\pi-\gamma$ system, one can integrate the
double differential width (spin-independent and spin-dependent) at
fixed values of variable $t,$ using the restrictions (7). For the
decay width $d\Gamma_0$ we have
\begin{equation}\label{55}
\frac{d\Gamma_0}{d\,t}=P\big[I_0(t)+(|a(t)|^2+|v(t)|^2)A_0(t)+Re(a(t))B_0(t)+Re(v(t))C_0(t)\big]\,,
\ P=\frac{Z^2}{2^8\pi^3M^2}\,,
\end{equation}
where $I_0(t)$ is the contribution of the inner bremsstrahlung
$$I_0(t)=\frac{4M}{t-m^2}\bigg\{\frac{M^2-t}{t}\big[(t+m^2)^2-4M^2t\big]+\big[2M^2(M^2+t-m^2)-m^4-t^2\big]L\bigg\}\,, \ L=\ln\frac{M^2}{t}\,.$$

As one can see, from Eq.~(55),\, the structure-dependent (resonance) contribution into $d\Gamma_0/dt$ does not contain
vector-axial-vector interference, but it includes a sum of the
squared moduli of the vector and axial-vector form factors. This
sum is multiplied by the function
$$A_0(t)=\frac{(t-m^2)^3(M^2-t)^2(M^2+2\,t)}{3M^5t^2}\,.$$

The interference of the IB and structural amplitudes includes only
real parts of the form factors and
$$B_0(t)=\frac{4(t-m^2)}{Mt}\big[(2t+M^2-m^2)(M^2-t)+t(m^2-2M^2-t)L\big]\,,$$
$$C_0(t)=\frac{4(t-m^2)^2}{Mt}\big(t-M^2+tL\big)\,.$$

As concerns the quantity $d\Gamma_1/dt$ connected with the Stokes
parameter $\xi_1$ (see Eq.~14 and formulas just below), it reads
\begin{equation}\label{56}
\frac{d\Gamma_1}{d\,t}=P\big[Im(a(t)^*v(t))A_1(t)+Im(v(t))C_1(t)\big]\,,
\end{equation}
where
$$A_1(t)=-2\frac{(t-m^2)^3(M^2-t)^3}{3M^5t^2}\,, \ \ C_1(t)=\frac{4(t-m^2)}{Mt}\big(t^2-M^4+2M^2t\,L\big)\,.$$
This quantity, in the case of an unpolarized $\tau,$ is the only one
which includes the imaginary parts of the vector-axial-vector
interference and imaginary part of the vector form factor. Because
it does not contain the pure IB contribution, it may be useful to
study the resonance one.

We have also
\begin{equation}\label{57}
\frac{d\Gamma_2}{d\,t}=P\big[I_2(t)+Re(a(t)^*v(t))A_2(t)+Re(a(t))B_2(t)+Re(v(t))C_2(t)\big]\,,
\end{equation}
 where
$$I_2(t)=-\frac{4M}{t}\big[(3t+m^2)(M^2-t)-t(t+m^2+2M^2)L\big]\,, $$
$$A_2=-2\,A_0(t)\,, \
B_2(t)=-C_0(t)\,, \ \ C_2(t)=-B_0(t)\,$$ for the part
corresponding to the circular polarization of the gamma quantum
(parameter $\xi_2$) and
\begin{equation}\label{58}
\frac{d\Gamma_3}{d\,t}=P\big[I_3(t)+\big(|a(t)|^2-|v(t)|^2\big)A_3(t)+Re(a(t))B_3(t)\big]\,,
\end{equation}
$$I_3(t)=\frac{8M(M^2-m^2)}{t-m^2}\big(2(t-M^2)+(M^2+t)L\big)\,,$$
$$A_3(t)=-\frac{1}{2}A_1(t)\,, \ \ B_3(t)=-C_1(t)\,$$
for the part connected with parameter $\xi_3\,.$

Obtained results are illustrated in Fig.~4 where we show the
quantities $d\,\Gamma_0(t)/d\,t$ and $\xi_i(t)$ defined as
$$\xi_i(t)=\frac{d\,\Gamma_i(t)/d\,t}{d\,\Gamma_0(t)/d\,t}\,, \ \ i=1\,, 2\,, 3$$

\vspace{0.4cm}

\begin{figure}
\captionstyle{flushleft}
\includegraphics[width=0.280\textwidth]{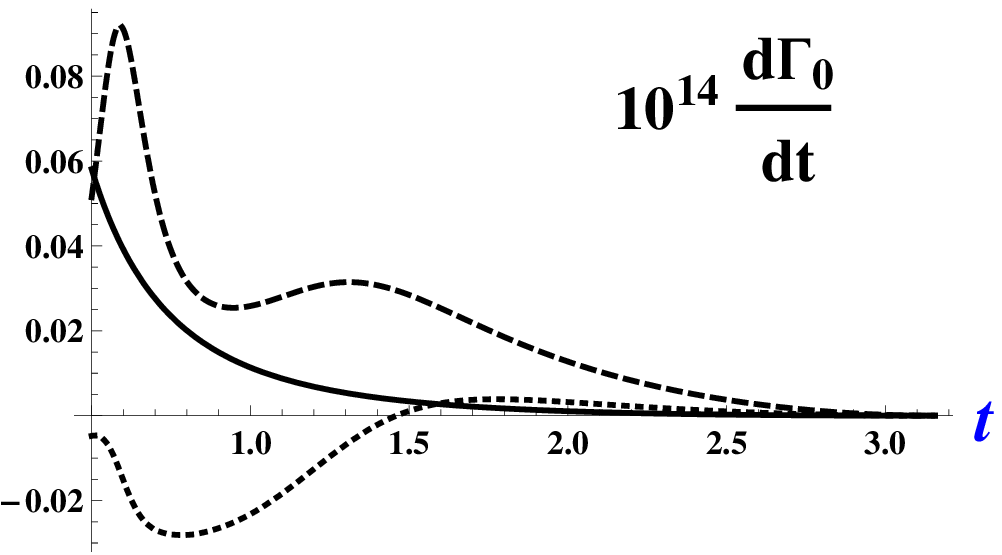}
\hspace{0.4cm}
\includegraphics[width=0.280\textwidth]{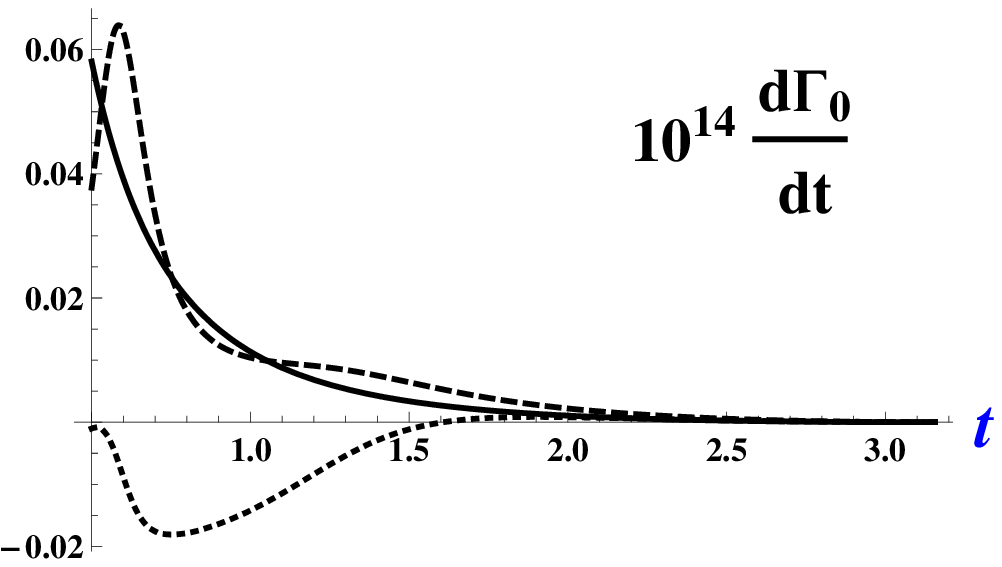}
\hspace{0.4cm}
\includegraphics[width=0.280\textwidth]{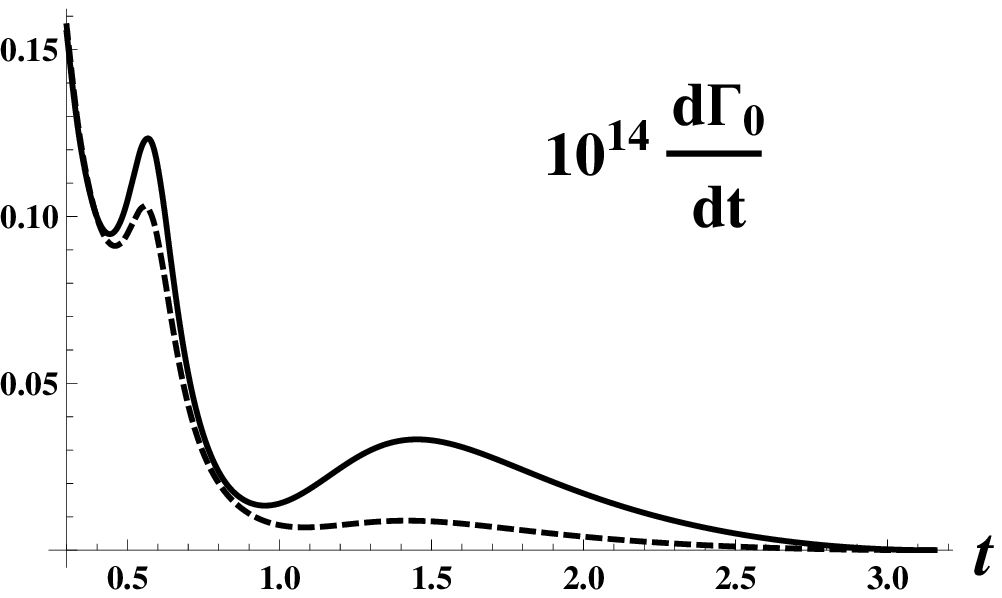}

\vspace{1.0cm}

\includegraphics[width=0.283\textwidth]{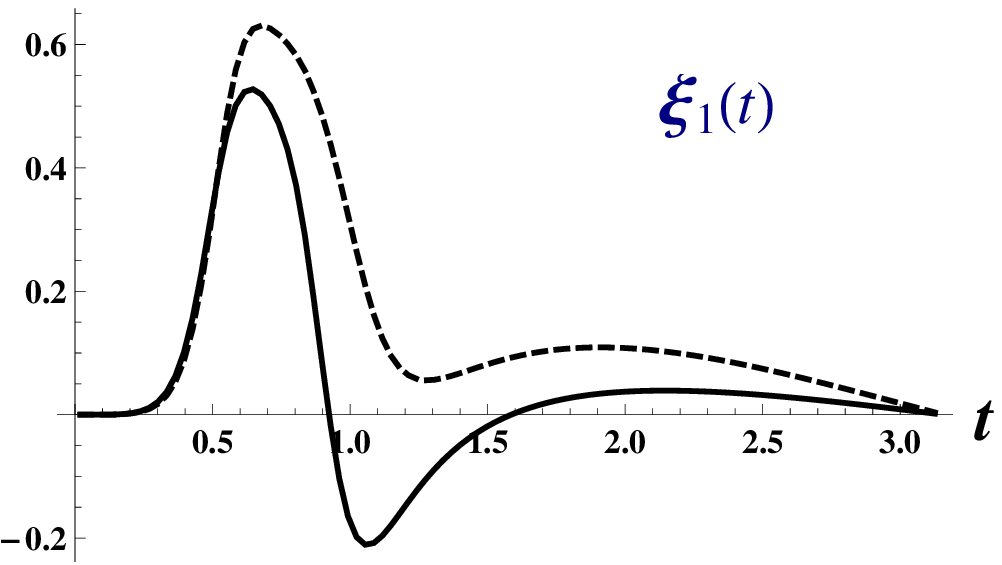}
\hspace{0.4cm}
\includegraphics[width=0.283\textwidth]{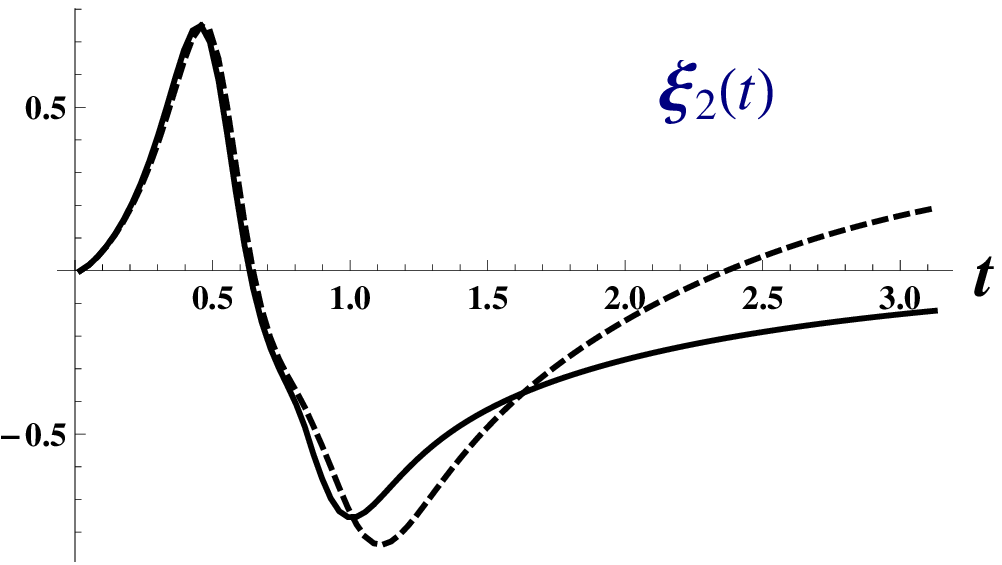}
\hspace{0.4cm}
\includegraphics[width=0.283\textwidth]{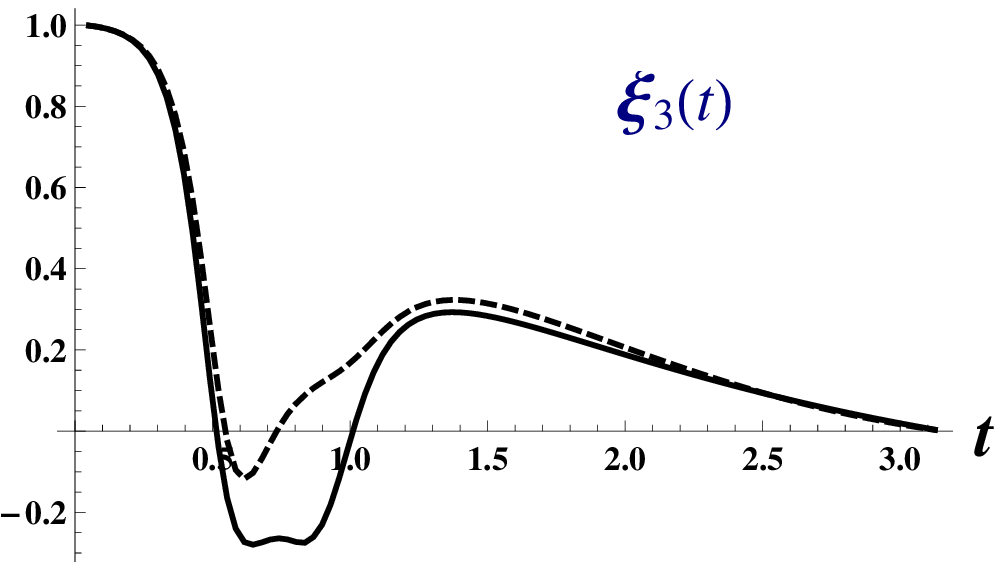}

\vspace{0.5cm}

\caption {The t-distribution of the
differential decay width, in GeV$^{-1},$ is shown in the upper
row. The first (second) figure corresponds to the set~1 (set~2) of
the resonance parameters given in the Table~1; the solid curve
represents the inner bremsstrahlung contribution, the dashed one -
resonance contribution and the dotted curve describes their
interference. The third figure shows the sum of all the
contributions: the solid (dashed) curve corresponds to the set~1
(set~2). The quantities $\xi_i,$ (see Eq.~(14)) in the lower row,
are calculated including all the contributions for two sets of the
parameters\,.}
\end{figure}

\vspace{0.5cm}

\subsection{The t-dependence for polarized $\tau^-$}

If $\tau^-,$ in the decay (1), is polarized, we can also write
analytical expressions for the quantities
$$\frac{d\Gamma^{^S}_0}{c_2\,dc_2\,dt}\,, \ \ \frac{d\Gamma^{^S}_i}{c_2\,dc_2\,dt}\,, \ i= 1\,, 2\,, 3\,.$$
%where the upper index $s$ indicate that in $\Sigma_i$ {\bf (see
%Eq.~(16))} only the parts depending on the $\tau^-$ polarization
%4-vector are taken into account.
They can be obtained from the corresponding fully differential
distributions by integration with respect to $c_1$ (using
relations (9)) and $\omega$ at fixed $t.$ Remind that in the rest
frame of $\tau^-,$ its polarization 3-vector is directed along the Z axis
and in this subsection we consider effects caused by the component
of this 3-vector which belongs to the decay plane
 $({\bf q\,,k}).$

The quantity, which defines the polarization asymmetry of the
decay, reads
\begin{equation}\label{59}
\frac{d\Gamma^{^S}_0}{c_2\,dc_2\,dt}=\frac{P}{2}\big[I^{^S}_0(t)+\big(|a(t)|^2+|v(t)|^2\big)A^{^S}_0(t)+
\end{equation}
$$Re(a(t)^*v(t))B^{^S}_0(t)+Re(a(t))C^{^S}_0(t)+Re(v(t))D^{^S}_0(t)\big]\,,$$
where
$$I^{^S}_0(t)=\frac{4\,M}{(t-m^2)}\big[-\frac{m^4M^2}{t}+m^4-2\,m^2M^2-6\,M^4+2\,m^2t+3\,M^2t+3\,t^2+$$
$$+[(m^2+M^2)^2+(M^2+t)^2+4\,M^2t]L\big]\,,$$
$$A^{^S}_0(t)=\frac{(t-m^2)^3}{3\,M^5t^2}\big[M^6-6M^4\,t+3M^2t^2+2\,t^3+6M^2t^2\,L\big]\,, \ \
B^{^S}_0(t)=\frac{4(t-m^2)^3}{M^3t}\big(M^2-t-t\,L\big)\,,$$
$$C^{^S}_0(t)=\frac{4(m^2-t)}{M\,t}\big[(t-M^2)(M^2+m^2+4\,t)+t(m^2+4M^2+t)L\big]\,,$$
$$D^{^S}_0(t)=\frac{4(m^2-t)}{M\,t}\big[(M^2-t)(m^2+3\,t)-t(m^2+2\,M^2+t)L\big]\,.$$

Quantities $d\Gamma^{^S}_i\,, \ i=1\,, 2\,, 3$ describe
correlations between the polarization states of $\tau^-$ and
photon. For them we have
\begin{equation}\label{60}
\frac{d\Gamma^{^S}_1}{c_2\,dc_2\,dt}=\frac{P}{2}\big[Im(a(t)^*v(t))B^{^S}_1(t)+Im(a(t))C^{^S}_1(t)+Im(v(t))D^{^S}_1(t)\big]\,,
\end{equation}
$$ B^{^S}_1(t)=A_1(t)\,, \ C^{^S}_1(t)=\frac{8(t-m^2)}{M}\big[2(M^2-t)-(M^2+t)L\big]\,,$$
$$D^{^S}_1(t)=\frac{4(t-m^2)}{M\,t}\big[(t-M^2)(5\,t+M^2)+2\,t(2M^2+t)L\big]\,;$$
\begin{equation}\label{61}
\frac{d\Gamma^{^S}_2}{c_2\,dc_2\,dt}=\frac{P}{2}\big[I^{^S}_2(t)+(|a(t)|^2+|v(t)|^2\big)A^{^S}_2(t)
+Re(a(t)^*v(t))B^{^S}_2(t)+
\end{equation}
$$Re(a(t))C^{^S}_2(t)+Re(v(t))D^{^S}_2(t)\big]\,,$$
$$I^{^S}_2(t)=-\frac{M^2}{t-m^2}D^{^S}_0(t)$$
$$A^{^S}_2(t)=\frac{2(t-m^2)^3}{M^3t}\big(t-M^2+t\,L\big)\,, \ \ B^{^S}_2(t)=-2A^{^S}_0(t)\,,$$
$$C^{^S}_2(t)=-D^{^S}_0(t)\,, \ D^{^S}_2(t)=-C^{^S}_0(t)\,;$$
\begin{equation}\label{62}
\frac{d\Gamma^{^S}_3}{c_2\,dc_2\,dt}=\frac{P}{2}\big[I^{^S}_3(t)+(|a(t)|^2-|v(t)|^2\big)A^{^S}_3(t)+Re(a(t))C^{^S}_3(t)+Re(v(t))D^{^S}_3(t)\big]\,,
\end{equation}
$$I^{^S}_3(t)=\frac{8\,M}{(t-m^2)t}\big[-(M^2-t)(t+2\,m^2+3\,M^2)+(M^4+3\,M^2\,t+m^2(t+M^2))L\big]\,,$$
$$A^{^S}_3(t)=-\frac{1}{2}A_1(t)\,, \ \ C^{^S}_3(t)=-D^{^S}_1(t)\,, \ D^{^S}_3(t)=-C^{^S}_1(t)\,.$$

It is easy to see that all the quantities and $d\Gamma^{S}_i\,$
vanish during the integration over the full angular region
 because they are proportional to $c_2\,d\,c_2$.

\vspace{0.4cm}

\begin{figure}
\captionstyle{flushleft}
\includegraphics[width=0.310\textwidth]{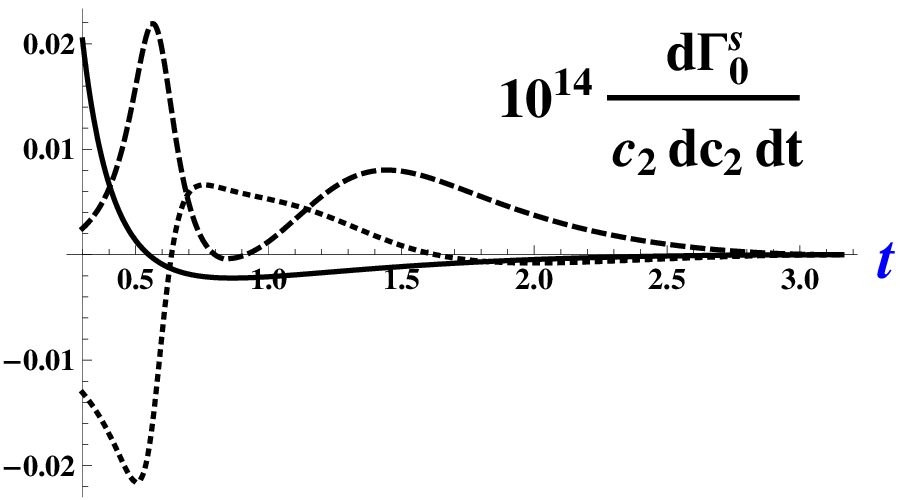}
\hspace{0.4cm}
\includegraphics[width=0.280\textwidth]{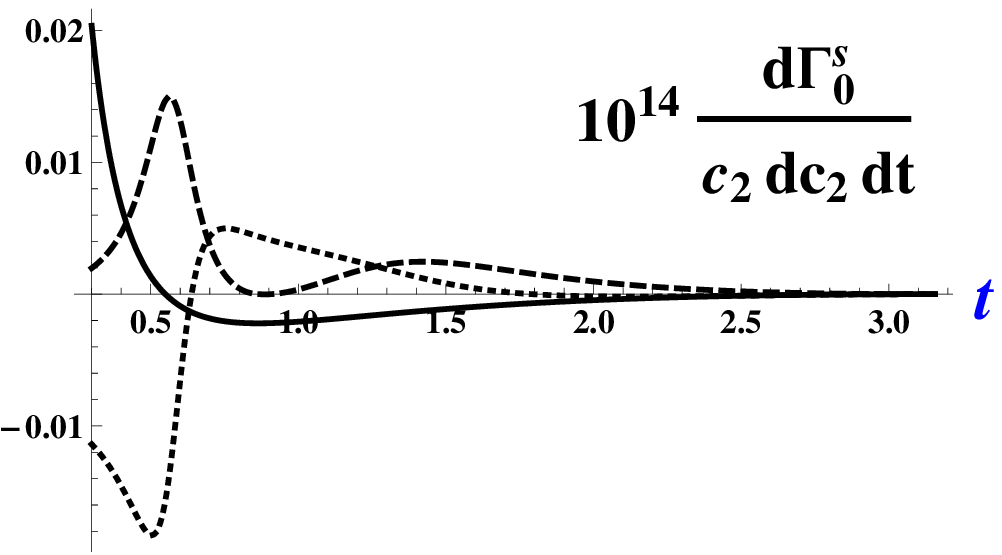}
\hspace{0.4cm}
\includegraphics[width=0.280\textwidth]{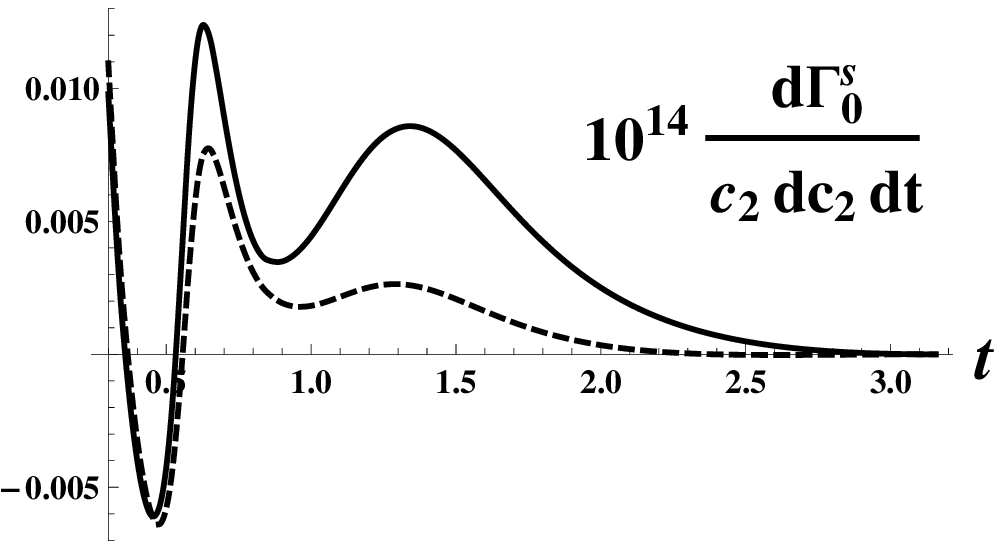}
\includegraphics[width=0.280\textwidth]{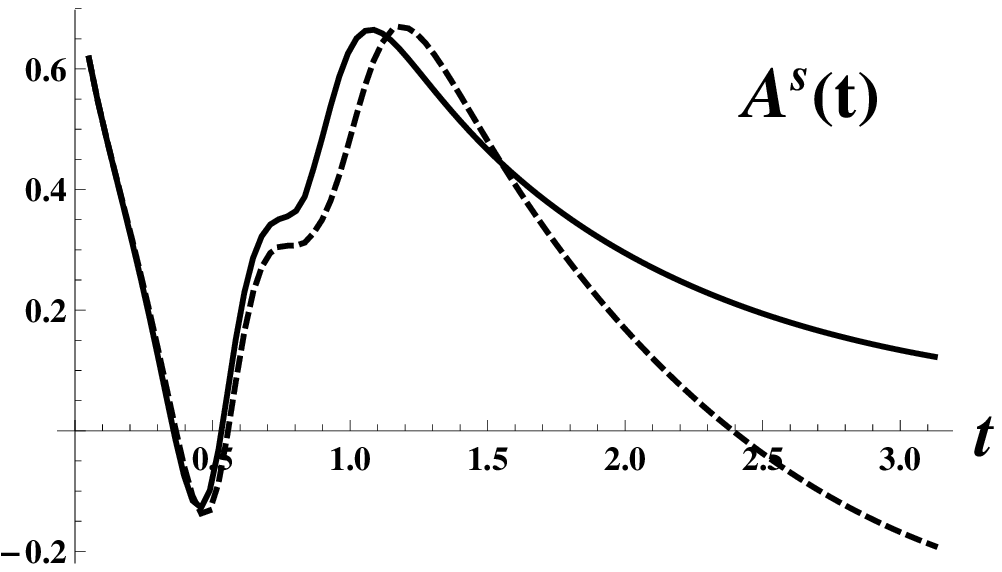}

\vspace{1.5cm}
\includegraphics[width=0.293\textwidth]{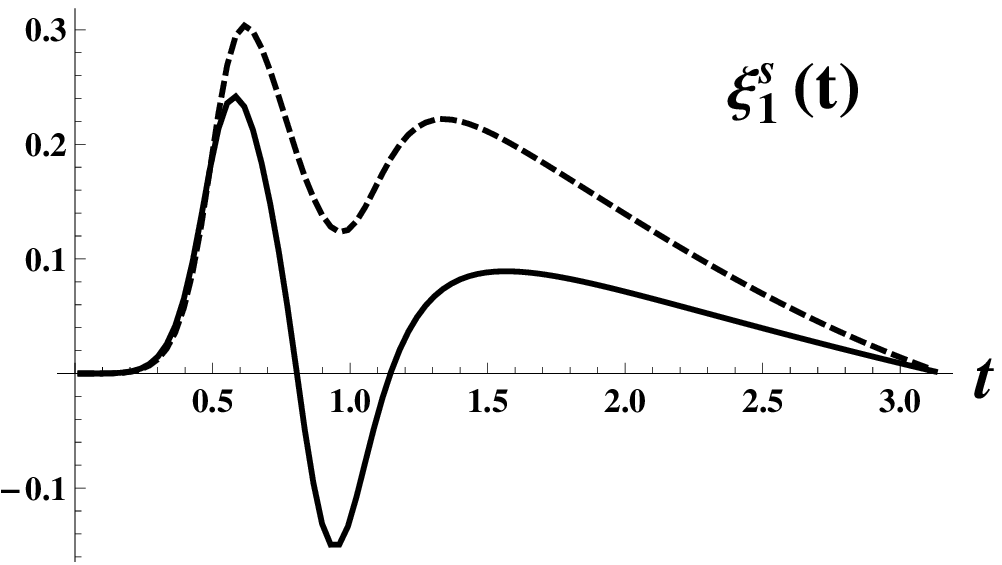}
\hspace{0.4cm}
\includegraphics[width=0.293\textwidth]{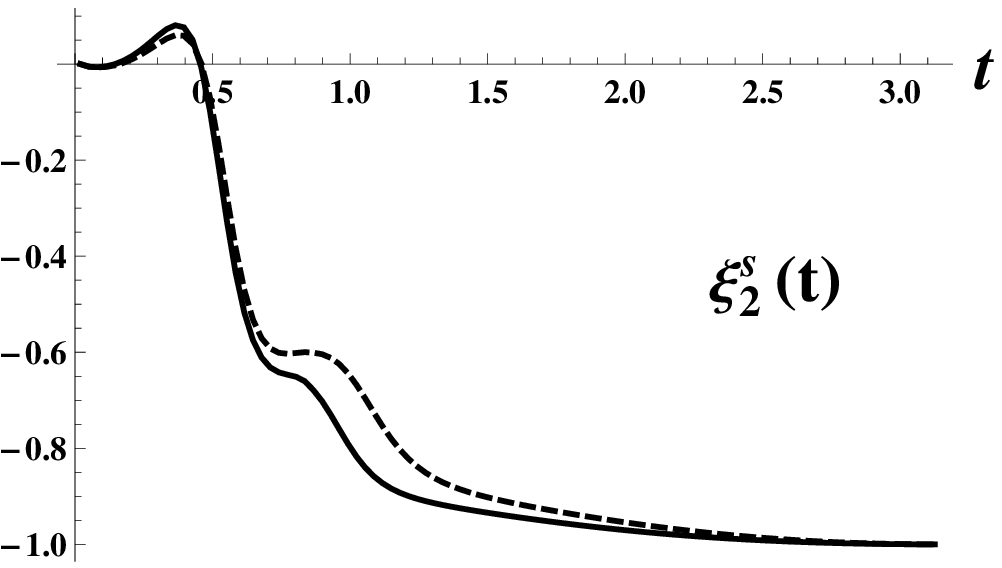}
\hspace{0.4cm}
\includegraphics[width=0.293\textwidth]{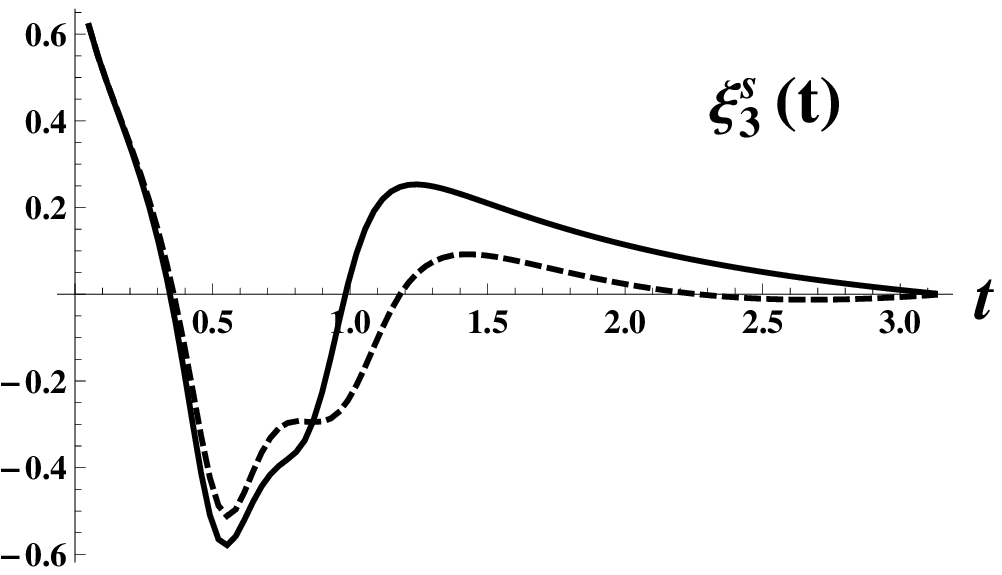}

\caption{The notation of the curves in the figures in the upper and lower
rows is the same as in Fig.~4. The figure in the middle row shows
the quantity defined in accordance with Eq.~(63) for two sets of
the parameters\,. Quantities $\xi^{^S}_i$ are defined in
Eq.~(64)\,.}

\end{figure}

In Fig.~5 we show the quantities: $d\Gamma^{^S}_0/c_2\,dc_2\,dt$
that is the part of the differential decay width which depends on
the $\tau^-$ polarization, the ratio
\begin{equation}\label{63}
A^{^S}(t)=\frac{2d\Gamma^{^S}_0/(c_2\,dc_2\,dt)}{d\Gamma_0/\,dt}
\end{equation}
about which we can say that $c_2A^{^S}(t)$ is the decay
polarization asymmetry at fixed values of $c_2$ and $t\,,$ as well
as the parameters
\begin{equation}\label{64}
\xi^{^S}_i(t)=\frac{2d\Gamma^{^S}_i/(c_2\,dc_2\,dt)}{d\Gamma_0/dt}\,,
\ \ i=1\,, 2\,, 3\,,
\end{equation}
which characterize different correlations between the
polarization states of $\tau^-$ and $\gamma$ quantum  in the
process (1).

\subsection{Dependence on the  photon energy}

The photon energy distribution requires integration over the
pion energy in the limits defined by the inequality (6). This
integration cannot be performed analytically because of non
trivial dependence of the vector and axial-vector form factors on
the pion energy. In this section we illustrate the results of our
numerical calculations for both unpolarized (Fig.~6) and polarized
(Fig.~7) $\tau$ lepton.

In Figs.~6 and 7 we restrict the photon
energy for the widths because at smaller energies
the IB contribution becomes very large (due to infrared divergence)
as compared with the structure-dependent one and their interference,
and the corresponding part of the figures is not illustrative.
As for the Stokes parameters, the expressiveness of the figures remains
good for the small energies.

\vspace{0.4cm}

\begin{figure}
\captionstyle{flushleft}
\includegraphics[width=0.260\textwidth]{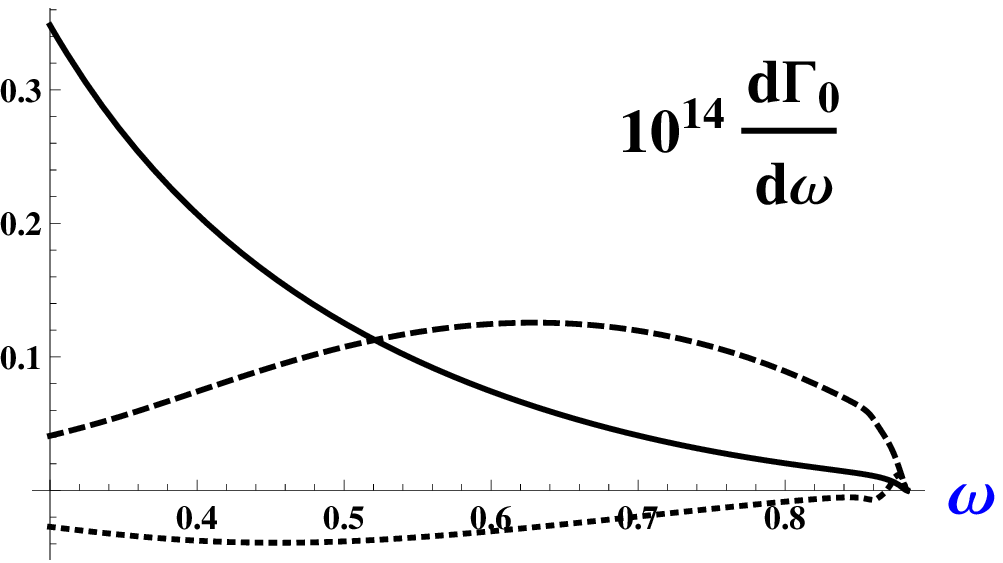}
\hspace{0.4cm}
\includegraphics[width=0.260\textwidth]{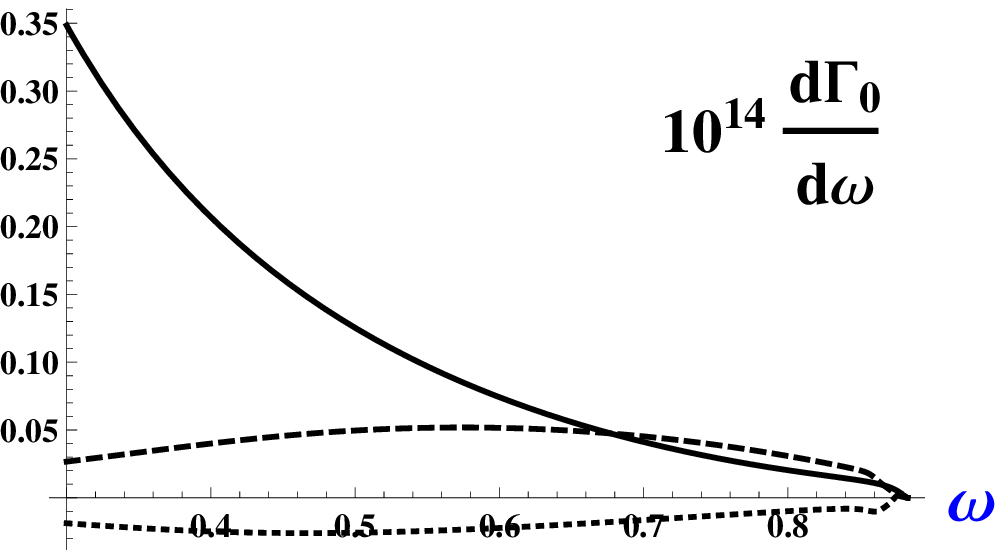}
\hspace{0.4cm}
\includegraphics[width=0.260\textwidth]{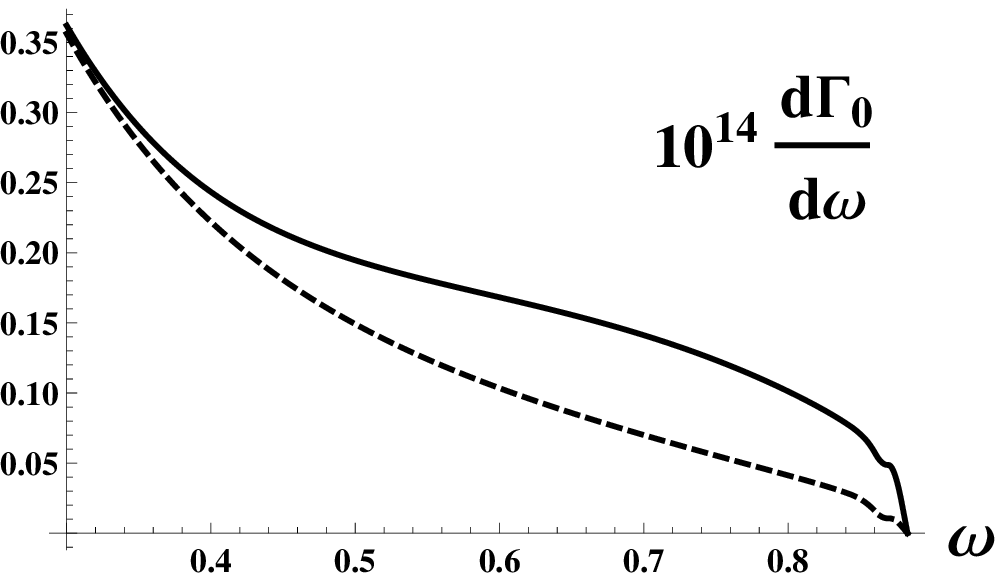}

\vspace{1.5cm}
\includegraphics[width=0.27\textwidth]{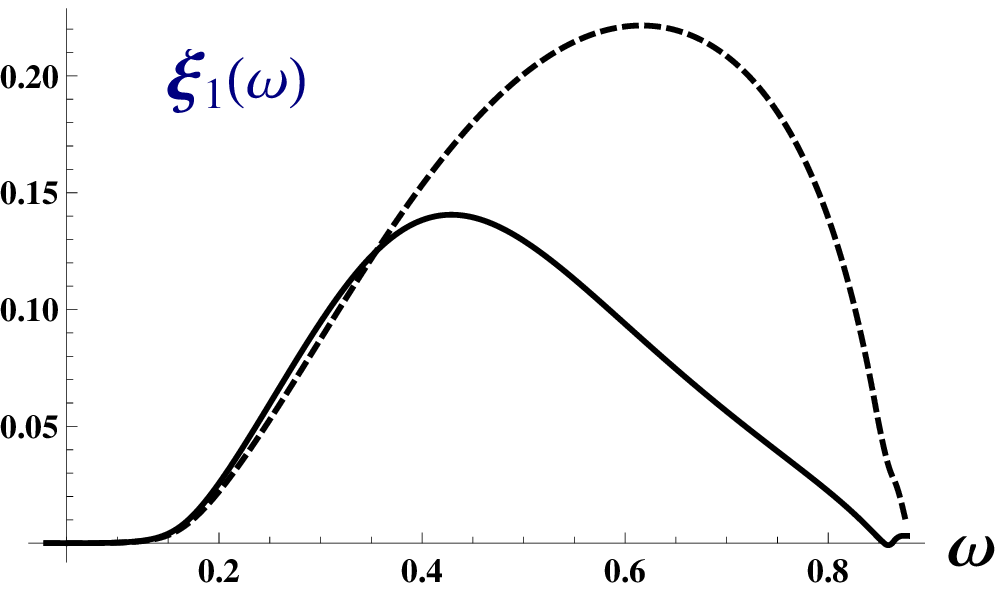}
\hspace{0.4cm}
\includegraphics[width=0.27\textwidth]{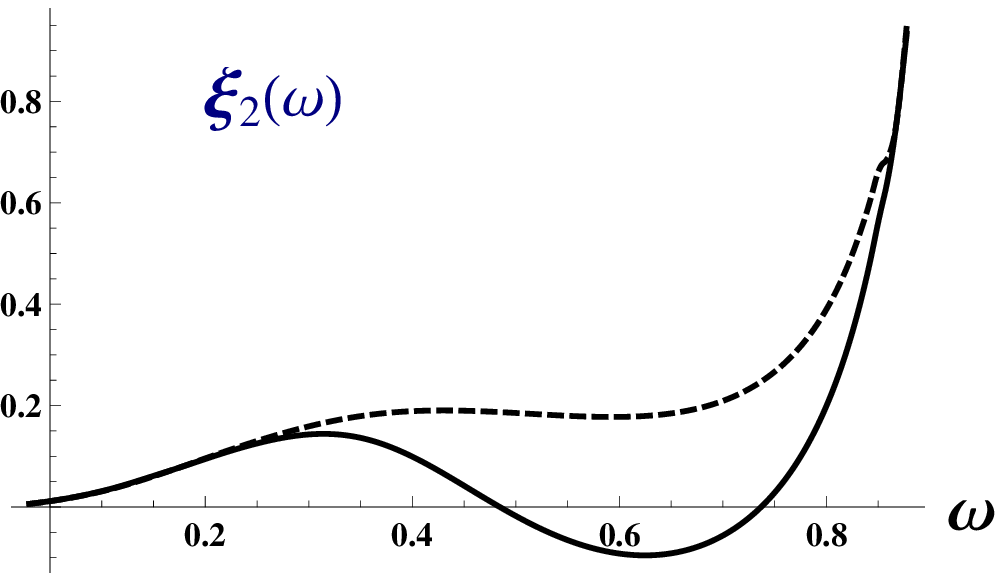}
\hspace{0.4cm}
\includegraphics[width=0.27\textwidth]{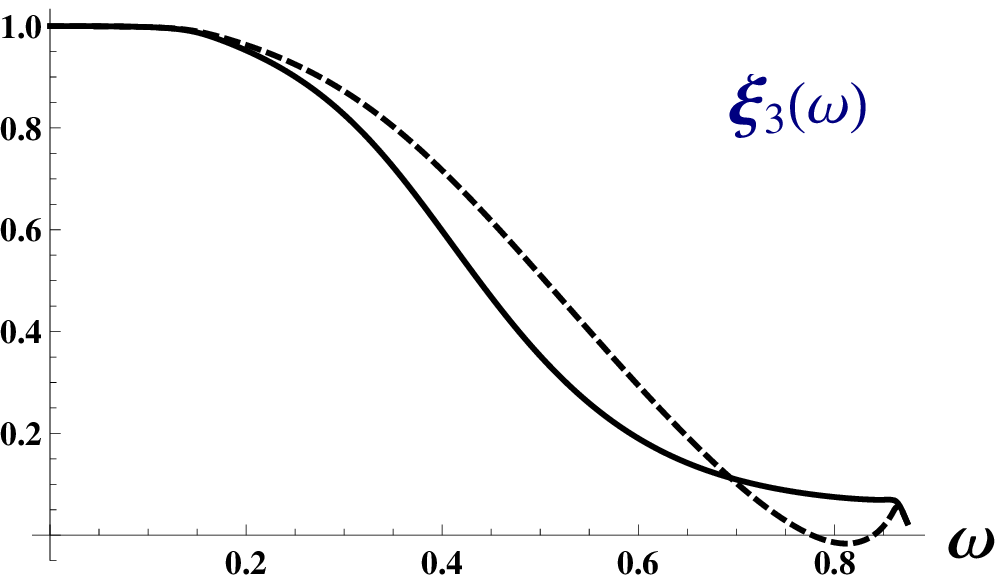}

\caption{The photon spectrum in the decay (1)
and the Stokes parameters versus the photon energy. Notation of
the curves is the same as in Fig.~4.}
\end{figure}

In the polarized case we define the quantities $A^{^S}(\omega)$ and
$\xi_i(\omega)$ in full analogy with the relations (63) and (64)
for $A^{^S}(t)$ and $\xi_i(t).$

\vspace{0.4cm}

\begin{figure}
\captionstyle{flushleft}
\includegraphics[width=0.310\textwidth]{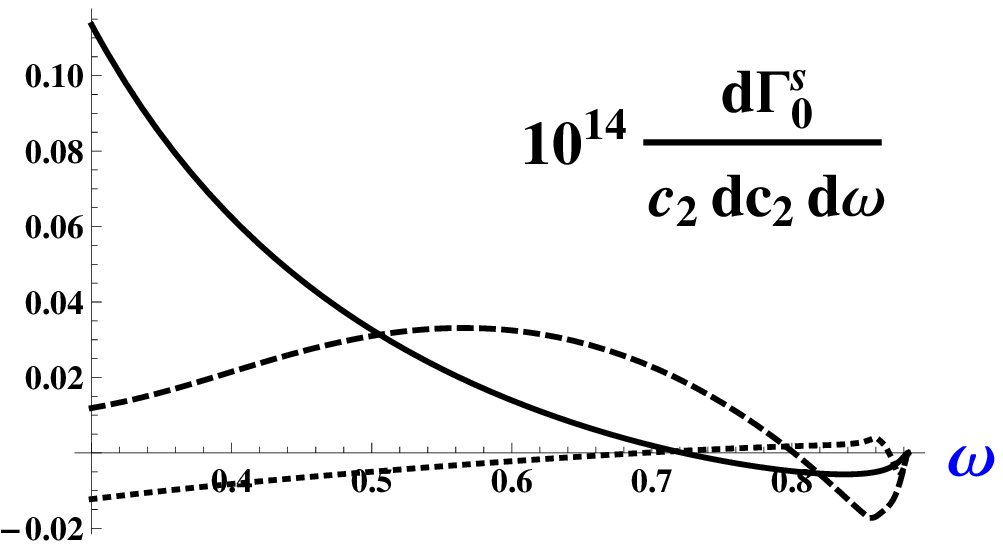}
\hspace{0.4cm}
\includegraphics[width=0.280\textwidth]{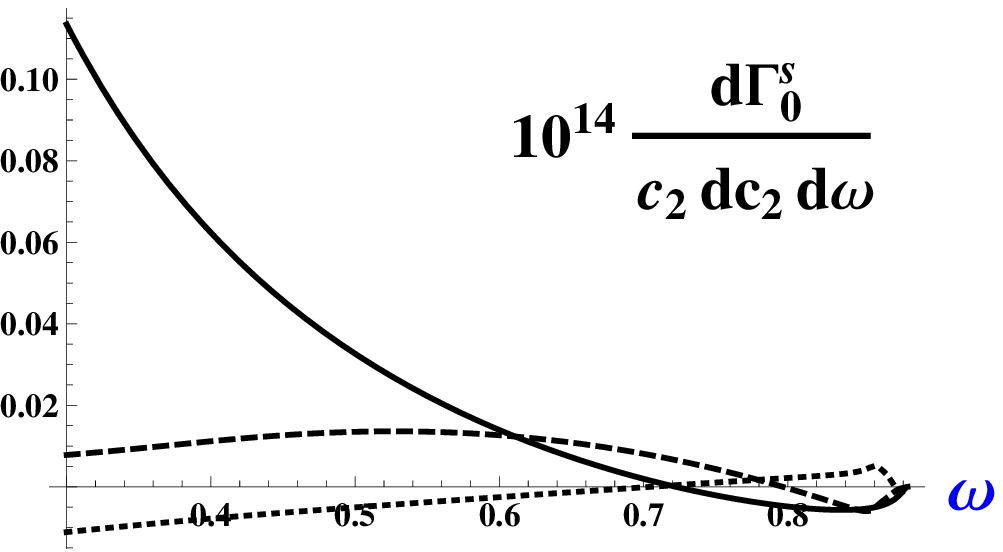}
\hspace{0.4cm}
\includegraphics[width=0.280\textwidth]{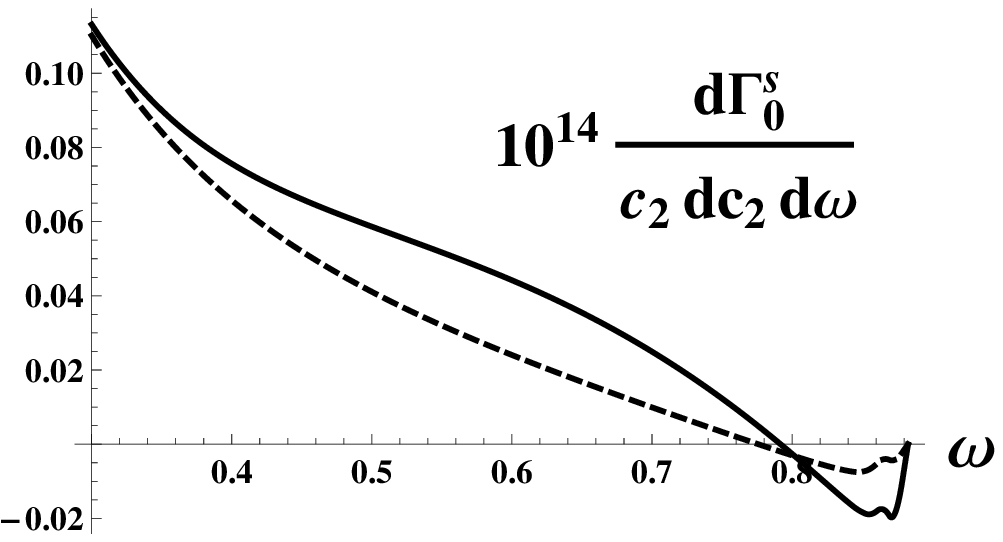}

\vspace{0.5cm}
\includegraphics[width=0.280\textwidth]{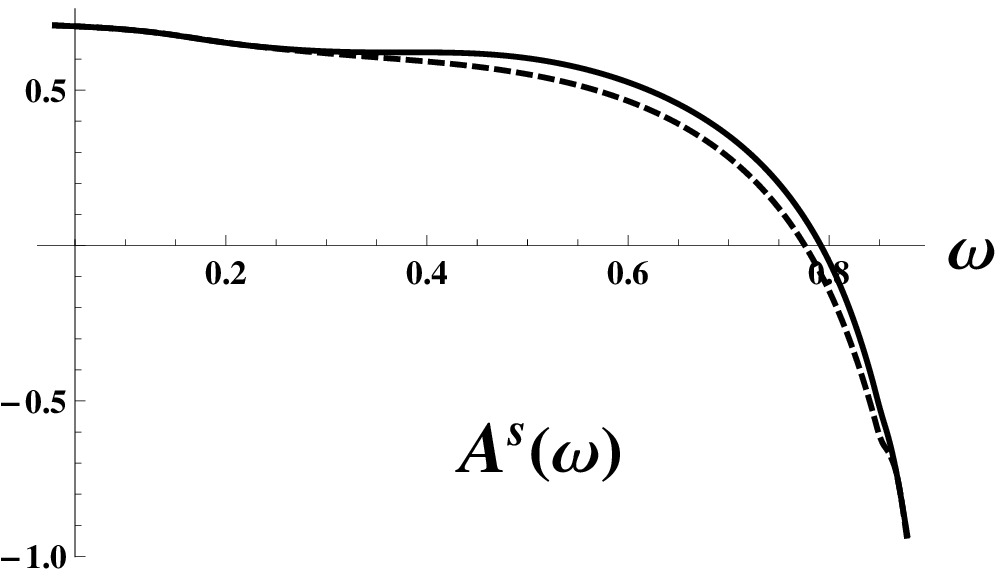}

\vspace{0.5cm}
\includegraphics[width=0.293\textwidth]{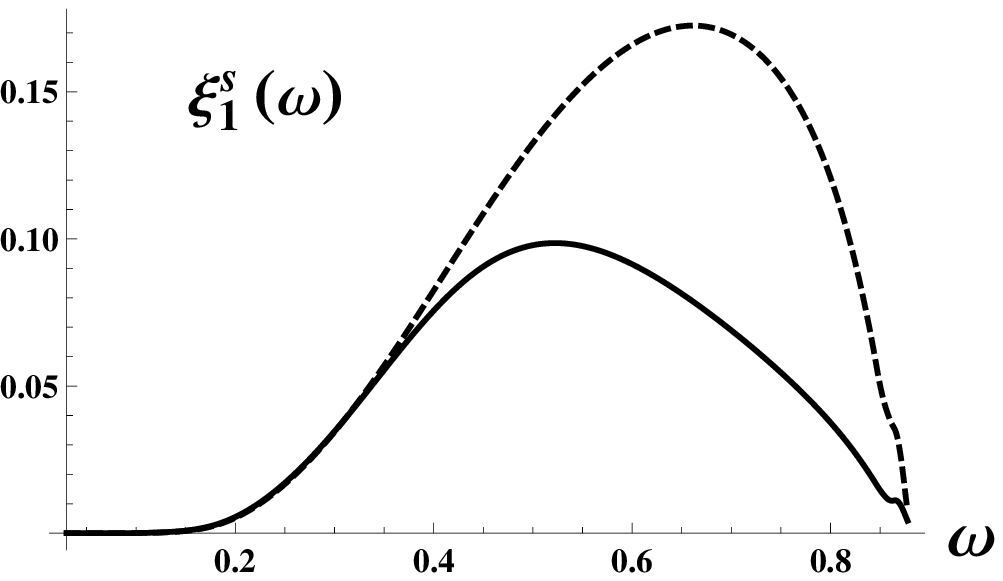}
\hspace{0.4cm}
\includegraphics[width=0.293\textwidth]{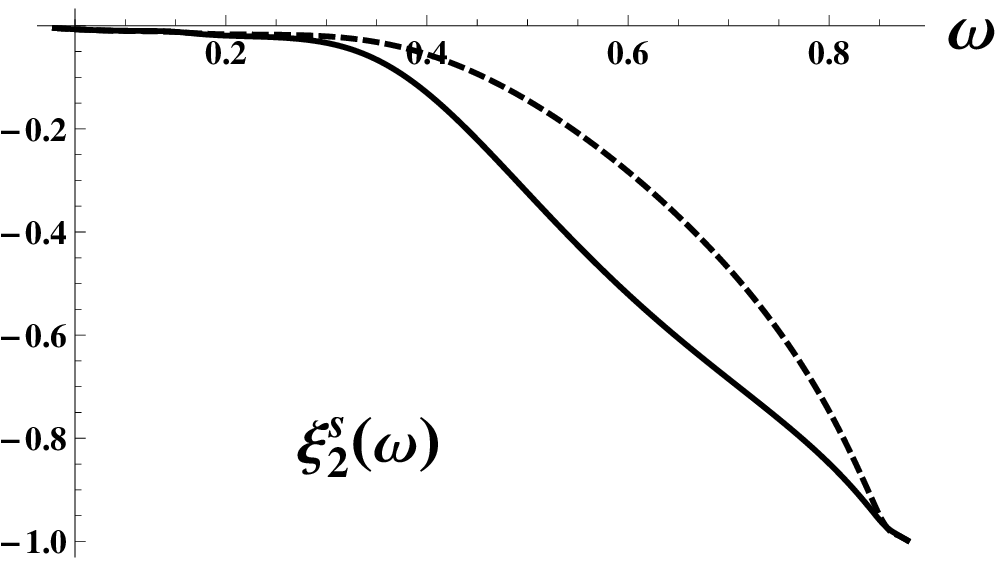}
\hspace{0.4cm}
\includegraphics[width=0.293\textwidth]{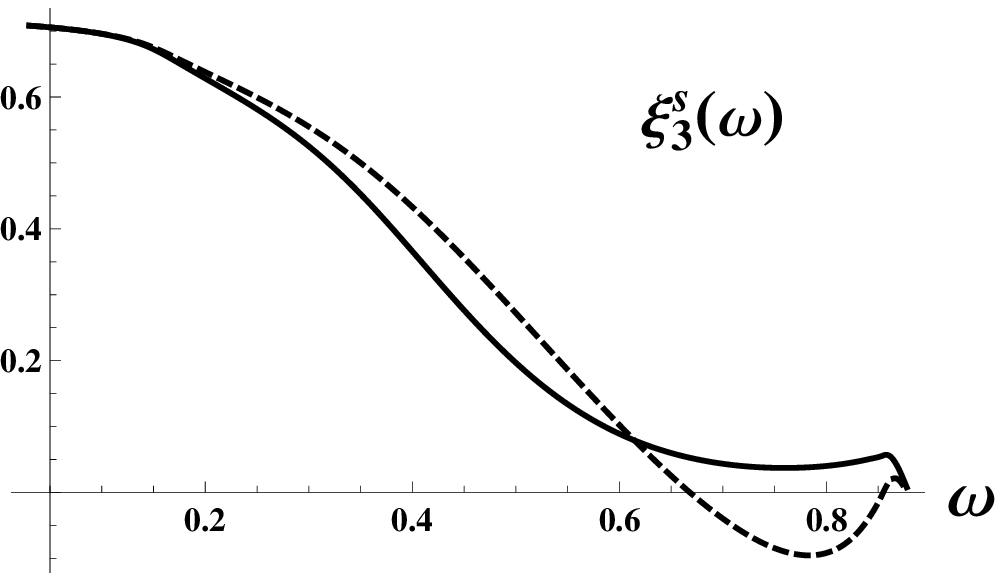}

\caption{The spin-dependent part of the photon
spectrum, the decay polarization asymmetry and the
spin-correlation parameters as a function of the photon energy.
Notation of the curves is the same as in Fig.~5. }
\end{figure}

\vspace{0.5cm}

\section{Discussion}

To determine the moduli and phases of the form factors, a procedure was
suggested in Ref. \cite{R95} (see Eq.~(66)) that does not require the
polarization measurements. For separation of contributions of
different form factor combinations it was suggested to measure
photon energy dependence of the differential probability $d\Gamma
/d\omega dt$ at fixed t value (i.e., at a fixed value of the sum of the
photon and pion energies).The obtained expression (in zero pion mass
approximation) for this quantity is a sum of the terms multiplied by
the photon energy to negative and positive powers. The measurement
of this distribution permits, in principle, to find the coefficients
in this power series. The following combinations of the form factors
$Rev(t)$, $Rea(t)$, $|v(t)|^2+|v(t)|^2$ and $Rea(t)v^*(t)$ can be
determined from these coefficients. Note that our calculations are performed  without
neglecting the pion mass. This is important for the decay $\tau^-\to
K^-\gamma\nu_{\tau}$ where it is necessary to take into account the
kaon mass.

Eq. (56) shows that the Stokes parameter $\xi_1$, as a function of
the variable t, is determined by the imaginary parts of the vector
and axial-vector form factors. But it follows from the unitarity
condition that $Imv(t)\neq 0$ for $t>4m^2$ and $Ima(t)\neq 0$ for
$t>9m^2$. So, this parameter must be zero for $t<4m^2$. Thus, the
value of the parameter $\xi_1(t)$ is completely determined by the
resonance contribution to the matrix element. The measurement of
this parameter in the region $t>4m^2$ can test the validity of this
mechanism for the description of the decay (1). Of particular interest is
region of the high values of the t variable where it
may be necessary to take into account the contribution of the
additional resonances beyond the $\rho$ and $a_1$ mesons which are
included in this paper. From Fig. 4 one can see that $\xi_1(t)$
parameter is sensitive to the choice of the parameters describing
the resonance contribution. In the region $1 GeV^2<t<1.5 GeV^2$, the
parameter $\xi_1(t)$ has opposite signs for the parameter set1 and
set2. The same conclusions are valid for the spin correlation
coefficient $\xi^s_1(t)$ (the Stokes parameter $\xi_1(t)$ which
depends on the $\tau$ lepton polarization vector), as it is seen
from Fig. 5. The Stokes parameters $\xi_1(\omega )$ and
$\xi^s_1(\omega)$, as a function of the photon energy, is also
sensitive to the choice of the parameter sets but in the region
$t>1.5 GeV^2$. In this case the signs of the parameters
$\xi_1(\omega )$ and $\xi^s_1(\omega)$ are the same for both
parameter sets.

The Stokes parameter $\xi_3(t)$ contains the contributions of the
IB, the interference between IB and resonance terms (which is
determined by the $Rea(t)$) and resonance term which depends on the
combination $|a(t)|^2-|v(t)|^2$. So, this parameter is less
sensitive to the choice of the parameter sets than the Stokes
parameter $\xi_1(t)$. Nevertheless, the sizeable sensitivity exists
in the region 0.5 GeV$^2$<t<1 GeV$^2$. The Stokes parameter
$\xi^s_3(t)$ is appreciably less sensitive to the choice of the
parameter sets than the parameter $\xi_3(t)$ (Fig. 5). The Stokes
parameter $\xi_3(\omega)(\xi^s_3(\omega))$, as a function of the
photon energy, is also less sensitive than the parameter
$\xi_1(\omega)(\xi^s_1(\omega))$.

Remind that the meaning of the parameters $\xi_1$ and $\xi_3$
requires the knowledge of the photon polarization vectors $e_1$ and $e_2.$
But to do this it is necessary to know the photon and pion momenta.
As to the parameter $\xi_2,$ in this case it is sufficient to know only photon momentum.

The Stokes parameter $\xi_2(t)$ contains the contributions of the
IB, the interference between IB and resonance terms (which is
determined by the $Rea(t)$ and $Rev(t)$) and resonance term which
depends on the $Re(a(t)v^*(t))$. From Fig. 4 one can see that this
parameter is sensitive to the choice of the parameter sets in the
region of the high values of the variable t (t>1 GeV$^2$). The
Stokes parameter $\xi^s_2(t)$ is weakly sensitive to the choice. The
corresponding parameters, as a function of the photon energy, show the
greater sensitivity to this choice in comparison with the same
parameters as functions of the t variable.

The photon energy has to be large enough to study the
sensitivity to the choice of the model parameters. Although the number of events in this
region is an order of magnitude smaller than in the low energy
photon one, where the IB-contribution dominates, one can expect that high
statistics precision measurements at the Super $c-\tau$ and SuperKEKB
factories make it possible to improve some model resonance
parameters used in our and in a number of other papers.

Note also that in Refs.~\cite{R95, Rekalo_1995} the authors suggested to study the
resonance mechanism in the radiative $\tau$ decay  by selecting
events (in the rest frame) at the maximal possible pion energy
$\varepsilon = (M^2+m^2)/(2M),$ where the contribution of IB to
the decay width vanishes due to the radiative zeros of electromagnetic
amplitudes for point-like particles \cite{EZ}. But the
corresponding  number of events decreases due to the essential
decrease of the kinematic region. On the other hand, one can
be sure that at chosen direction of axes the IB contribution to the
Stokes parameter $\xi_1(\varepsilon\,, ~ \omega)$ vanishes in the
whole kinematic region.

Most of the analytic calculations presented in this paper can also be used  for analysis of the decay
$\tau^- \to \nu_\tau K^- \gamma$. All the results in sections 2 and 3 will remain the same,
apart from trivial changes of the masses, form factors and CKM matrix elements. Moreover, the
$t$-distributions, obtained in section 5, will retain their form, but will be expressed in terms of
the kaon mass and the corresponding vector and axial-vector form factors. The latter can be
derived in framework of the R$\chi$T following the procedure in section 4. We plan to perform
calculations for the decay $\tau^- \to \nu_\tau K^- \gamma$ in the future.

\section{Conclusion}

We have investigated the radiative one-meson decay of the $\tau $
lepton, $\tau^-\to \pi^-\gamma\nu_{\tau}$. The photon energy
spectrum and t-distribution (t is the square of the invariant mass
of the pion-photon system) of the unpolarized $\tau $
lepton decay have been calculated. We have also studied the polarization
effects in this decay. The following polarization observables have
been calculated: the asymmetry caused by the $\tau $ lepton
polarization, the Stokes parameters of the emitted photon and spin
correlation coefficients which describe the influence of the $\tau
$ lepton polarization on the photon Stokes parameters.

The amplitude of the $\tau $ lepton decay, $\tau^-\to
\pi^-\gamma\nu_{\tau}$, has two contributions: the inner
bremsstrahlung which does not contain any free parameters and
structure-dependent term which is parameterized in terms of the
vector and axial-vector form factors. Note that in our case these
form factors are the functions of the t variable and t$>0$, i.e.,
we are in the time-like region. The form factors in this region
are the complex functions and their full determination, i. e.,
not only their moduli but their phases as well, is non-trivial in
this case. To do this it is necessary to perform the polarization
measurements.

We calculated the unpolarized and polarized observables for two
sets of the parameters describing the vector and axial-vector form
factors. The numerical estimation shows that some polarization
observables (the asymmetry and the Stokes parameters, especially
$\xi_1$) can be effectively used for the discrimination between
two parameter sets since these observables have opposite signs in
some regions of the variable t or the photon energy.

\begin{center}
{\bf Acknowledgement}
\end{center}

This work was supported by the National Academy of Sciences of
Ukraine under PICS No. 5419.

\begin{appendix}

\section*{Appendix A. Interactions in framework of the chiral theory with resonances \label{app:A}}
 \setcounter{equation}{0}
\def\theequation{A.\arabic{equation}}
 \label{sec:weak-em-interaction}

In this Appendix we outline the framework for a calculation of the
form factors for the decay $\tau^-  \to \nu_\tau \pi^- \gamma $.
We use the formalism of the chiral theory with resonances
(R$\chi$T) suggested in Refs.~\cite{EckerNP321, EckerPL223}.  The
corresponding Lagrangian can be written as
\begin{equation}
 {\cal L}_{R \chi T} \, = \,  {\cal L}_{\chi}^{(2)}  \, + \, {\cal L}_R \, ,
 \label{eq:Lagrangian}
\end{equation}
where  the  $SU(3)_L \otimes SU(3)_R$ chiral Lagrangian in the
order ${\cal O}(p^2)$ is
\begin{equation}
{\cal L}_{\chi}^{(2)} \, = \, \frac{F^2}{4} \, \langle u_{\mu} \,
u^{\mu} \, + \, \chi_{+} \rangle \,, \label{eq:L^2}
\end{equation}
 \begin{eqnarray}
\label{eq:uchi}
 u_{\mu} & = & i [ u^{\dagger}(\partial_{\mu}-i r_{\mu})u-
u(\partial_{\mu}-i l_{\mu})u^{\dagger} ] \ , \nonumber \\
\chi_{+} & = & u^{\dagger}\chi u^{\dagger} + u\chi^{\dagger} u\, ,
\quad \quad   \chi=2B_0(s+ip) \, ,
\end{eqnarray}
where $\langle \ldots \rangle$ means the trace in the flavor space
and $F$  is the pion weak decay constant in the chiral limit.

The octet of the pseudoscalar mesons $P$ with $J^P=0^-$ is
included in the matrix $$ u (\Phi) = \exp (i\Phi /\sqrt{2} F) = 1
+ i\Phi /\sqrt{2} F - \Phi^2 / 4 F^2 + ... ,$$
 where $\Phi$ is
\bge \Phi \, = \,  \left(
\begin{array}{lcr}
\pi^0/\sqrt{2} +\eta_8/\sqrt{6} & \pi^+ & K^+ \\
\pi^-         & - \pi^0/\sqrt{2} +\eta_8/\sqrt{6} & K^0 \\
K^-  & \bar{K}^0  & -2 \eta_8 /\sqrt{6}
\end{array}  \right)\, .
 \ene

The masses of the pseudoscalar mesons enter into
Eq.~(\ref{eq:L^2}) via the quark mass matrix ${\cal M}$
 \bgea
\chi =  2 B_0 {\cal M} + ...,  \qquad \qquad  {\cal M}={\rm
diag}(m_u, \, m_d,\, m_s) \,  ,
 \label{eq:mass}
 \enea
and the constant $B_0$ is expressed through  the quark condensate
$ \langle \bar{q} q  \rangle$
 \bgea
 B_0 = - \frac{ \langle \bar{q} q \rangle}{F^2} = \frac{M_{\pi^{0, \pm}}^2}{m_u + m_d} =
 \frac{M_{K^0}^2}{m_d + m_s} = \frac{M_{K^\pm}^2}{m_u + m_s}= \frac{3 M_\eta^2}{m_u + m_d + 4m_s}.
 \label{eq:mass-1}
 \enea
The value $ \langle \bar{q} q \rangle \approx (-240 \pm 10 \ {\rm
MeV})^3 \; \; ({\rm at \ an \ energy \ scale} \; \mu=1 \; {\rm
GeV}) $.  In the limit of exact isospin symmetry \ $ \chi \,= {\rm
diag} (m_\pi^2, m_\pi^2, 2m_K^2 - m_\pi^2)$ in terms of the masses
of $\pi$ meson, $m_\pi$, and $K$ meson, $m_K$.

The interaction of the pseudoscalar mesons with the $W^\pm_\mu$, \
$Z_\mu$-bosons and electromagnetic field $B_\mu$ can be included
via the
%vector, $v_{\mu} = (r_{\mu} + l_{\mu})/2$, and axial-vector, $a_{\mu} =
%(r_{\mu} - l_{\mu})/2$,
external fields $r_{\mu}$ and $l_{\mu}$ as follows
 \bgea
r_\mu   &=& - e Q B_\mu + g \frac{\sin^2 \theta_W}{\cos \theta_W}
Q  Z_\mu,
\nn \\
l_\mu &=& - e Q B_\mu - \frac{g}{2 \sqrt{2}} \, W_\mu  +
\frac{g}{\cos \theta_W} (Q \sin^2 \theta_W + \frac{1}{6} - Q)Z_\mu
\,. \label{eq:right_left_fields}
   \enea
Here the quark charge matrix is $Q = {\rm
diag}(\frac{2}{3},-\frac{1}{3},-\frac{1}{3})$, \  $e = \sqrt{4 \pi
\alpha}$ is the positron charge, $g=e/ \sin \theta_W$ is the
$SU(2)_L$ coupling constant,  and $\theta_W$ is the weak angle.

We also introduced the notation
\begin{equation}
 W_\mu \, \equiv \, W_\mu^+ T_+ +  W_\mu^- T_- \, ,
\end{equation}
with $W_\mu^\pm= (W_1 \mp W_2)_\mu$ being the field of the charged
weak bosons, and matrices $T_+, \, T_-$ are defined as
\begin{equation}
\label{eq:T}
 T_+  = \left(
\begin{array}{ccc}
0  & V_{ud} &  V_{us} \\
0 &  0 & 0  \\
 0  &  0 & 0
\end{array}
\right) \ ,  \qquad \qquad  T_- = T_+^\dagger \,, \nonumber
\end{equation}
where  $V_{ud} $ and $V_{us}$ are the elements of the CKM
matrix~\cite{CKM}.

The lowest-order even-intrinsic-parity Lagrangian ${\cal L}_R$,
describing interaction of the resonance fields with the
pseudoscalars, has been suggested in Ref.~\cite{EckerNP321}. It is
linear in the resonance fields. In this Lagrangian we keep the
contributions from the vector and axial-vector mesons relevant for
the process of $\tau$  decay to $\pi$ and $\gamma$:
\begin{eqnarray}
\label{eq:lagr-R-inter} {\cal L}_{R} &=& {\cal L}_{kin}^R + {\cal
L}_{int}^R,
\\
{\cal L}_{kin}^R &=& - \frac{1}{2} \sum_{R=V,A}\left\langle
\nabla^\lambda R_{\lambda \mu}\, \nabla_\nu R^{\nu \mu}
 - \frac{M_R^2}{2}  R_{\nu \mu} R^{\nu \mu} \right\rangle ,
 \nonumber \\
{\cal L}_{int}^R &=& \frac{F_{V}}{2\sqrt{2}} \left\langle V_{\mu
\nu } f_+ ^{\mu \nu}\right\rangle + \frac{ iG_{V}}{\sqrt{2}}
\left\langle V_{\mu \nu }u^{\mu }u^{\nu } \right\rangle + \frac{
iF_{A}}{2 \sqrt{2}} \left\langle A_{\mu \nu } f_- ^{\mu \nu}
\right\rangle\, , \nonumber
\end{eqnarray}
where $F_V, \, G_V, \, F_A$ are the coupling constants, and
antisymmetric tensor representation for the spin-1 fields $V_{\mu
\nu}$ and $A_{\mu \nu}$ is applied \cite{EckerNP321}.

Also here
 \bgea f^{\mu\nu}_{\pm} = u F^{\mu\nu}_L u^\dagger \pm u^\dagger
F^{\mu\nu}_R u, &\ \ \ \ \ & F^{\mu\nu}_R = \partial^\mu r^\nu -
\partial^\nu r^\mu - i [r^\mu, r^\nu],
\\
\nn &\ \ \ \ \ & F^{\mu\nu}_L = \partial^\mu l^\nu - \partial^\nu
l^\mu - i [l^\mu, l^\nu],
\\
\nn \nabla_\mu X = \partial_\mu X + [\Gamma_\mu, X],
 &\ \ \ \ \ &
\Gamma_\mu   = \sfrac{1}{2} \ \{ u^\dagger (\partial_\mu - i
r_\mu) u + u (\partial_\mu - i l_\mu) u^\dagger \}.
 \enea
The fields $R=\{ V,A\}$ represent nonets (octet and singlet) of
the vector and axial-vector resonances with the lowest masses. In
general, the higher-lying multiplets can be added if needed.
However, already the low-lying resonances saturate the low-energy
constants of $\chi$PT~\cite{EckerNP321}. We do not include here
interactions nonlinear in the resonance fields; the structure of
such terms has been investigated in
Refs.~\cite{Gomez_2004,Ruiz-Femenia_2003,Cirigliano_2006}.

For description of the vector form factor one needs the
interaction Lagrangian in the odd-intrinsic-parity sector. The
lowest order ${\cal O}(p^4)$ interactions follow from the
Wess-Zumino-Witten (WZW) functional~\cite{WZW}. It is sufficient
to keep there only the terms which are linear in the pseudoscalar
fields $\Phi$ and containing two external fields $l_\mu, \,
r_\mu$. Thus we retain
\begin{eqnarray}
{\cal L}^{(2)}_{WZW} &=& - \frac{\sqrt{2} N_C}{48 \pi^2 F}\,
\epsilon^{\mu\nu\rho\sigma} \, {\rm Tr}\,
 \Big[ \partial_\mu \Phi \, \Big( \frac{1}{2} l_\sigma\partial_\nu r_\rho  +
 \frac{1}{2}  r_\sigma  \partial_\nu l_\rho + \frac{1}{2} r_\nu \partial_\rho  l_\sigma
+ \frac{1}{2} l_\nu \partial_\rho  r_\sigma + \nonumber \\
&& + l_\nu
\partial_\rho l_\sigma  + r_\nu \partial_\rho  r_\sigma +
l_\sigma\partial_\nu l_\rho + r_\sigma  \partial_\nu r_\rho
 \Big) \Big]\,
 \label{eq:WZW_1}
\end{eqnarray}
where $N_C =3$ is the number of the quark colors, and $\epsilon^{0
1 2 3} = +1$.

The external fields $r_\mu, \, l_\mu$ are further expressed
through the electromagnetic field and the field of the $W$ boson
in Eq.~(\ref{eq:right_left_fields}). Choosing in
Eq.~(\ref{eq:WZW_1}) only the electromagnetic field one would
obtain $\pi^0 \gamma \gamma$, \ $\eta \gamma \gamma $ and
$\eta^\prime \gamma \gamma$ interactions. Here we are interested
in the terms proportional to both fields, of the photon and
$W$-boson:
\begin{eqnarray}
{\cal L}_{\gamma W \Phi} &=&  \frac{3 e g N_C}{48 \pi^2 F}\,
\epsilon^{\mu\nu\rho\sigma} \, {\rm Tr}\,
 \Big[ \Phi \, \Big( \partial_\mu B_\nu Q  \partial_\rho W_\sigma
 + \partial_\rho W_\sigma  Q \partial_\mu B_\nu
 \Big) \Big]
\label{eq:WZW_2}  \\
&=&  \frac{e g N_C}{48 \pi^2 F}\, \epsilon^{\mu\nu\rho\sigma}
\partial_\mu B_\nu \Big[ V_{ud} (\pi^- \partial_\rho W_\sigma^+   + \pi^+ \partial_\rho W_\sigma^-)
+ V_{us} (K^- \partial_\rho W_\sigma^+   + K^+ \partial_\rho
W_\sigma^-) \Big]\,. \nonumber
\end{eqnarray}
An additional odd-intrinsic-parity interaction relevant for the
transition $W \to V \to P \gamma $ is considered in Section 4.

From the Lagrangians (\ref{eq:Lagrangian}), (\ref{eq:WZW_1}) and
(\ref{eq:WZW_2}) one can obtain the necessary terms describing
interactions of  the pseudoscalar mesons, resonances, $W$ boson
and photon.

\end{appendix}

%%%%%%%%%%%%%%%  Appendix B   %%%%%%%%%%%%%%%%%%

%%%%%%%%%%%%%%%%%%%%%%  Appendix C %%%%%%%%%%%%%%%%%%%%%%%%%%%%%

\begin{appendix}
\section*{Appendix B. Polarization of $\tau^-$ lepton in electron-positron annihilation}
\label{app:B} \setcounter{equation}{0}
\def\theequation{B.\arabic{equation}}

Let us consider the polarization of the $\tau^-$ lepton   in
annihilation process
$$e^-(p_1)+e^+(p_2)\rightarrow \tau^-(q_1)+\tau^+(q_2) $$
provided the electron has nonzero longitudinal polarization.
The effect can be understood by means of the corresponding matrix
element squared

$$|M|^2= E^{\mu\nu}T_{\mu\nu}\ , \
\frac{1}{2}E^{\mu\nu}=-\frac{s}{2}g^{\mu\nu}+(p_1p_2)^{\mu\nu}+i\lambda(\mu\nu
p_1p_2)\ ,$$
\begin{equation}\label{B1}
\frac{1}{2}\ T_{\mu\nu}=
-\frac{s}{2}g^{\mu\nu}+(q_1q_2)^{\mu\nu}+iM(\mu\nu qS)\ , \
q=p_1+p_2=q_1+q_2\ ,\ q^2=s\ ,
\end{equation}
where $S$ is the $\tau$ polarization 4-vector,
%$c=\cos{\theta}$
and $\lambda$ is the electron polarization degree.

The contraction of the tensors in Eq.~(B1) reads (neglecting the
electron mass)
\begin{equation}\label{B2}
|M|^2= A+B(S)\ , \
A=s^2\Big[1+\cos^2{\theta}+\frac{4M^2}{s}\sin^2{\theta}\Big]\,,  \
B(S)= -4\lambda Ms[(p_1S)-(p_2S)]\,,
\end{equation}
where $\theta$ is the angle between the momenta of the electron
and $\tau^-$ lepton.

The polarizations of $\tau^-$ are defined as follows \cite{Berest}
\begin{equation}\label{B3}
P^L=\frac{B(S^L)}{A}\ , \ \ P^T=\frac{B(S^T)}{A}\,.
\end{equation}
If we choose the 4-vectors  $S^L$ and $S^T$ in such a way that in
the rest frame of the $\tau^-$ lepton they are
\begin{equation}\label{B4}
S^L=(0,{\bf n_1})\ , \ \ \ S^T=(0,{\bf n_2})\ , \ \ {\bf n_1\bf
n_2}=0\ ,\ \ {\bf n_1^2=\bf n_2^2}=1\ ,
\end{equation}
where $\bf n_1$ is in the direction of the 3-vector $\bf q_1$ in
c.m.s. and $\bf n_2$ belongs to the scattering plane, then we can
write the covariant form  of the 4-vectors $S^L$ and $S^T$ in
terms of the momentum 4-vectors, namely (neglecting the electron
mass)
$$ S^L=\frac{(q_1q_2)q_1-M^2q_2}{M\sqrt{(q_1q_2)^2-M^4}}\ , $$
\begin{equation}\label{B5}
S^T=\frac{[M^2-(p_2q_1)]q_1+[M^2-(p_1q_1)]q_2+[(p_1p_2)-2M^2]p_1}{N}\
,
\end{equation}
$$N^2=[(p_1p_2)-2M^2][2(q_1p_1)(q_1p_2)-M^2(p_1p_2)]\ .$$
It is easy to show that in the rest frame of $\tau^-$ covariant
forms in Eq.~(B5) coincides with relations given in Eq.~(B4).
Using the above formulae one finds
\begin{equation}\label{B6}
B(S^L)=2\lambda s^2\cos{\theta}\ , \ \ B(S^T)=4\lambda M
s\sqrt{s}\sin{\theta}\
\end{equation}
and the corresponding results for the polarizations $P^L$ and
$P^T$ are given in the text.

At the Super $c - \tau$ factory planned in Novosibirsk,
$\tau$-pairs will be created near the threshold where  the
directions of their 3-momenta are not determined, therefore above
evaluations are not convenient. The only marked direction, for the
considered reaction in this case in $c.m.s.,$ is the colliding
beams one. It means that choosing in Eq.~(B4) the unit 3-vector
$\bf n_1$ along the electron beam direction ${\bf p}_1$ and, as
before, $\bf n_2$ is in the reaction plane, we are able to go to
the threshold limit and clear interpret the $\tau^-$ polarization
states.

This requires a modification of the corresponding covariant forms
for the polarization 4-vectors as follows
$$ S^l=\frac{1}{M}\bigg(\frac{M^2}{(p_1q_1)}p_1-q_1\bigg)\,, $$
\begin{equation}\label{B7}
S^t=\frac{1}{N_1}\bigg[\bigg(\frac{M^2}{(p_1q_1)}-1+\frac{(p_1q_1)}{(p_1p_2)}\bigg)p_1+
\frac{(p_1q_1)}{(p_1p_2)}p_2-q_1\bigg]\,,
\end{equation}
$$N_1^2=2(p_1q_1)-M^2-\frac{2(p_1q_1)^2}{(p_1p_2)}
\,.$$

In this case we have
$$B(S^{l,t})=2\lambda s^2D^{l,t}\,, \ \ D^l= \frac{4M^2}{s}
\bigg[1-\sqrt{1-\frac{4M^2}{s}}\cos{\theta}\bigg]^{-1}-\sqrt{1-\frac{4M^2}{s}}\cos{\theta}\,,
$$
$$D^t=-\frac{2M}{\sqrt{s}}\sqrt{1-\frac{4M^2}{s}}\sin{\theta}\,,$$
and near the threshold when
$$1-\frac{4M^2}{s}\ll 1$$
we have
\begin{equation}\label{B8}
P^l=\lambda\Big[1-\frac{1}{2}\Big(1-\frac{4M^2}{s}\Big)\sin^2{\theta}\Big]\,
\  \ P^t=-\frac{2\lambda
M\,\sin{\theta}}{\sqrt{s}}\sqrt{1-\frac{4M^2}{s}}\,.
\end{equation}
Just at the threshold $(s=4M^2)$ $P^l=\lambda, \ P^t=0.$

Thus we see that near the threshold the $\tau$ lepton practically
keeps the longitudinal polarization of the electron beam.
\end{appendix}


\begin{thebibliography}{99}
\bibitem{P13}
A. Pich, arXiv:hep-ph/1310.7922.
\bibitem{Z11}
M. Zobov, Part. Nucl. {\bf 42}, 1480 (2011).
\bibitem{K08}
K. Inami, Nucl. Phys. {\bf B 181-182} (Proc. Suppl.), 295 (2008).
\bibitem{Lev08}
E.~Levichev, Part. Nucl. Lett. {\bf 5}, 554 (2008).
\bibitem{O09}
K.~Oide, Prog. Theor. Phys. {\bf 122}, 69 (2009).
\bibitem{K80}
J. H. Kim, L. Resnick, Phys. Rev. D {\bf 21}, 1330 (1980).
\bibitem{B86}
S. Banerjee, Phys. Rev. D {\bf 34}, 2080 (1986).
\bibitem{I90}
Yu.P. Ivanov, A.A. Osipov, M.K. Volkov,
Phys. Lett. B {\bf 242}, 498 (1990).
\bibitem{D93}
R. Decker, M. Finkemeier, Phys. Rev. D {\bf 48}, 4203 (1993).
\bibitem{R95}
N. K. Pak, M. P. Rekalo, T. Yilmaz, Ann. Phys. {\bf 241}, 447 (1995).
\bibitem{G03}
C. Q. Geng, C. C. Lih, Phys. Rev. D {\bf 68}, 093001 (2003).
\bibitem{G10}
Zhi-Hui Guo, P. Roig, Phys. Rev. D {\bf 82}, 113016 (2010).
%\bibitem{R14}
%P. Roig, arXiv:hep-ph/1401.4219.
%\bibitem{G14}
%A. Guevara, arXiv:hep-ph/1401.6411.
\bibitem{Guevara:2013wwa}
  P. Roig, A. Guevara, and G. L$\acute{o}$pez Castro,
  %``Weak radiative pion vertex in $\tau^- \to \pi^-\nu_\tau \ell^+ \ell^-$ decays,''
  Phys.\ Rev.\ D {\bf 88}, no. 3, 033007 (2013).

\bibitem{E89}
S. Egli et al., Phys. Lett. B {\bf 222}, 533 (1989).
\bibitem{B92}
V. A. Baranov et al., Sov. Jour. Nucl. Phys. 55 (1992) 1644.
\bibitem{PDG}
J. Beringer $et \  al.$ (Particle Data Group), Phys. Rev. D {\bf
86}, 010001 (2012).
\bibitem{EckerNP321} G.~Ecker, J.~Gasser, A.~Pich, E.~de~Rafael,
Nucl.~Phys.~\textbf {B~321}, 311 (1989).
\bibitem{EckerPL223} G.~Ecker, J.~Gasser, H.~Leutwyler, A.~Pich,
E.~de~Rafael, Phys. Lett. B \textbf{223}, 425 (1989).

\bibitem{Cirigliano_2006} V. Cirigliano, G. Ecker,  M. Eidem\"{u}ller, R. Kaiser,  A. Pich, J. Portol\'{e}s, Nucl. Phys. {\bf B 753}, 139 (2006).

\bibitem{Rosell_2007} I.~Rosell,  PhD Thesis (IFIC, Departament de F\'{i}zica Te\`{o}rica),  e-Print: hep-ph/0701248.

\bibitem{Portoles_2010}  J.~Portol\'{e}s,   AIP Conf.\ Proc.\  {\bf 1322}, 178 (2010).

\bibitem{Dubinsky_2005} S. Dubinsky, A. Korchin, N. Merenkov, G. Pancheri, O. Shekhovtsova, Eur. Phys. J. C {\bf 40}, 41 (2005);
\ \  S.~Eidelman, S.~Ivashyn, A.~Korchin, G.~Pancheri and
O.~Shekhovtsova, Eur. Phys. J. C {\bf 69}, 103 (2010).

\bibitem{Ivashyn_2008} S.A.~Ivashyn and A.Yu.~Korchin,  Eur. Phys. J. C {\bf 54}, 89 (2008).

\bibitem{Czyz_2012} K.~Kampf and J.~Novotny,
  %``Resonance saturation in the odd-intrinsic parity sector of low-energy QCD,''
  Phys.\ Rev.\ D {\bf 84}, 014036 (2011); \\
H.~Czy\.{z}, S.~Ivashyn,  A.~Korchin and O.~Shekhovtsova, Phys. Rev. D {\bf 85}, 094010 (2012); \\
P.~Roig, A.~Guevara and G.~L.~Castro,
  %``The VV'P form factors in resonance chiral theory and the pi-eta-eta' light-by-light contribution to the muon g-2,''
  Phys.\ Rev.\ D {\bf 89}, 073016 (2014).

\bibitem{Nugent:2013hxa}  I.~M.~Nugent, T.~Przedzinski, P.~Roig, O.~Shekhovtsova and Z.~Was,
  %``Resonance chiral Lagrangian currents and experimental data for $\tau^-\to\pi^{-}\pi^{-}\pi^{+}\nu_{\tau}$,''
  Phys.\ Rev.\ D {\bf 88}, 093012 (2013).

\bibitem{RekaloLuca}M.P. Rekalo, Acta Physica Polonica\, B {\textbf 17}, 391 (1986)\,;
\ E. Gabrielli, L.Trentadue, Nucl. Phys. {\bf B 792}, 48 (2008).
\bibitem{CKM} N.~Cabibbo,  Phys.\ Rev.\ Lett. {\bf 10}, 531 (1963);
\ M.~Kobayashi and T.~Maskawa,  Prog.\ Theor.\ Phys.\ {\bf 49},
652 (1973).
\bibitem{Exptau}
A.~Lusiani, "The physics of the new facilities Super Flavour
Factories", Talk at 9th Franco-Italian Meeting on B Physics, 18-19
February 2013, LAPP.
\bibitem{Bl09}
A.~Blinov $et.\, al.$ ICFA Beam Dyn. Newslett. {\bf 48}, 268 (2009).
\bibitem{Berest} V.B. Berestetskii, E.M. Lifshitz, L.P. Pitaevskii,
Quantum Electrodynamics (in Russian), Moscow, Nauka, 1989, p. 291;
\ English translation: Quantum Electrodynamics. Volume 4 of Course
of Theoretical Physics, Second edition 1982. Pergamon Press.
Oxford - New York - Toronto - Sydney - Paris - Frankfurt, p. 652.

\bibitem{Rekalo_1995} N.K. Pak, M.P. Rekalo, T. Yilmaz, Ann. Phys. \textbf{241}, 416 (1995).

\bibitem{Ecker_2003} J.~H.~Kuhn and A.~Santamaria,   Z.\ Phys.\ C {\bf 48}, 445 (1990); \\
G. Ecker and R. Unterdorfer,  Nucl.~Phys.~Proc.~Suppl. {\bf 121}, 175 (2003).

 \bibitem{Dumm:2009va}
  D.~G.~Dumm, P.~Roig, A.~Pich and J.~Portoles,
  %``tau ---> pi pi pi nu(tau) decays and the a(1)(1260) off-shell width revisited,''
  Phys.\ Lett.\ B {\bf 685}, 158 (2010).

\bibitem{Gomez_2004} D. Gomez Dumm, A. Pich, and J. Portol\'{e}s, Phys. Rev. D {\bf 69}, 073002 (2004).

\bibitem{Ruiz-Femenia_2003} P.D. Ruiz-Femenia, A. Pich, and J. Portol\'{e}s, J. High Energy Phys. {\bf 07}, 003 (2003).

\bibitem{WZW} J.~Wess and B.~Zumino, Phys. Lett. B {\bf 37} (1971) 95; \ E.~Witten, Nucl. Phys. {\bf B 223}, 422 (1983).

\bibitem{Geng_2004} C.Q.~Geng, I-Lin~Ho, T.H.~Wu, Nucl.~Phys. {\bf B 684}, 281 (2004).

\bibitem{Unterdorfer_2008} R.~Unterdorfer, H.~Pichl, Eur.~Phys.~J. C {\bf 55}, 273 (2008).

%\bibitem{Bijnens_1997} J.~Bijnens, P.~Talavera, Nucl.~Phys. B489 (1997) 387-404.

\bibitem{EckerPLB237} G.~Ecker, A.~Pich, E.~de~Rafael, Phys. Lett. B {\bf 237}, 481 (1990).

%\bibitem{Prades} J.~Prades, Z. Phys. C \textbf{63} (1994) 491; \
%{\it Erratum}, Eur. Phys. J. C \textbf{11} (1999) 571.

\bibitem{Lepage} G.P. Lepage and S.J. Brodsky, Phys. Lett. B {\bf 87}, 359 (1979); \  S.J. Brodsky and G.P. Lepage, Phys. Rev. D {\bf 24},
1808 (1981).

\bibitem{Roig:2014}   P.~Roig and J.~J.~Sanz-Cillero,
  %``Consistent high-energy constraints in the anomalous QCD sector,''
  Phys.\ Lett.\ B {\bf 733}, 158 (2014)



\bibitem{EZ}
R.~Brown, D.~Sahdev, and K.~Mikaelan, Phys.Rev. D {\bf 20}, 1164
(1979); S.J.~Brodsky and R. W.~Brown, Phys.Rev.Lett. {\bf 49}, 966
(1982).
%\bibitem{SMM08}
%J. She, Y. Mao, B.Q. Ma, Phys. Lett. B {\bf 666} (2008) 355.
%\bibitem{PV05}
%B. Pausquini and M. Vanderhaeghen, Eur.Phys. J. A {\bf 24} (2005)
%29.
%\bibitem{AKM}
%A.V. Afanasev, M.I. Konchatnij, N.P. Merenkov, Zh. Eksp. Teor.
%Fiz. {\bf 129} (2006) 254.
%\bibitem{GRT05}
%G.I. Gakh, A.P. Rekalo, E. Tomasi-Gustafsson, Ann. of Phys. {\bf
%319} (2005) 150.


\end{thebibliography}
\end{document}